\documentclass[]{emulateapj}
\usepackage{color}

\usepackage{amsmath}
\usepackage{lscape}
\usepackage{ulem}
\usepackage{apjfonts}

\def\ltsima{$\; \buildrel < \over \sim \;$}
\def\simlt{\lower.5ex\hbox{\ltsima}}
\def\gtsima{$\; \buildrel > \over \sim \;$}
\def\simgt{\lower.5ex\hbox{\gtsima}}

\newcommand{\lx}{$L_{\rm X}$}
\newcommand{\nh}{$N_{\rm H}$}

\shorttitle{Comoving Space Density and Obscured Fraction of High-Redshift Active Galactic Nuclei in the Subaru/{\it XMM-Newton} Deep Survey}
\shortauthors{K. Hiroi, Y. Ueda, M. Akiyama, and M. G. Watson}

\begin{document}

\title{Comoving Space Density and Obscured Fraction of High-Redshift Active Galactic Nuclei in the Subaru/{\it XMM-Newton} Deep Survey}

\author{Kazuo HIROI\altaffilmark{1}, Yoshihiro UEDA\altaffilmark{1},
Masayuki AKIYAMA\altaffilmark{2}, and Mike G. WATSON\altaffilmark{3}}
\email{hiroi@kusastro.kyoto-u.ac.jp}

\altaffiltext{1}{Department of Astronomy, Kyoto University, Oiwake-cho, Sakyo-ku, Kyoto, Japan}
\altaffiltext{2}{Astronomical Institute, Graduate School of Science, Tohoku University, Aramaki, Aoba, Sendai, Japan}
\altaffiltext{3}{X-Ray and Observational Astronomy Group, Department of Physics and Astronomy, University of Leicester, University Road, Leicester LE1 7RH, UK}
\altaffiltext{*}{Based in part on data collected at Subaru Telescope, which is operated by the National Astronomical Observatory of Japan.}

\begin{abstract}

We study the comoving space density of X-ray-selected luminous active
galactic nuclei (AGNs) and the obscured AGN fraction at high redshifts
($3 < z < 5$) in the Subaru/{\it XMM-Newton} Deep Survey (SXDS) field. From
an X-ray source catalog with high completeness of optical identification 
thanks to deep optical images, we 
select a sample of 30 AGNs at $z > 3$ with intrinsic (de-absorbed and 
rest-frame 2--10 keV) luminosities of $L_{\rm X} = 10^{44-45}$ erg s$^{-1}$ detected in the 
0.5--2 keV band, consisting of 20 and 10 objects with 
spectroscopic and photometric redshifts, respectively. 
Utilizing the $1/V_{\rm max}$ method, we
confirm that the comoving space density of luminous AGNs decreases with
redshift above $z > 3$. When combined with the {\it Chandra}-COSMOS result of
Civano et al.\ (2011), the density decline of AGNs with $L_{\rm X} = 10^{44-45}$ 
erg s$^{-1}$ is well represented by a power law of $(1 +
z)^{-6.2 \pm 0.9}$. We also determine the fraction of X-ray obscured AGNs
with $N_{\rm H} > 10^{22}$ cm$^{-2}$ in the Compton-thin population to be
0.54$^{+0.17}_{-0.19}$, by carefully taking into account observational
biases including the effects of photon statistics for each
source. This result is consistent with an independent determination of the type-2
AGN fraction based on optical properties, for which the fraction is found to be
0.59$\pm$0.09. Comparing our result with that obtained in the local 
Universe, we conclude that the obscured fraction of luminous AGNs 
increases significantly from $z=0$ to $z>3$ by a factor of 
2.5$\pm$1.1.

\end{abstract}


\section{INTRODUCTION}

The evolution of active galactic nuclei (AGNs) carries key information
for understanding the growth history of supermassive black holes (SMBHs)
in galactic centers. Recent X-ray surveys have revealed that luminous
AGNs have their number density peak at higher redshifts, $z \sim 2-3$,
than less luminous ones \citep{ueda2003, barger2005, lafranca2005, 
hasinger2005, silverman2008, ebrero2009, yencho2009,
aird2010}. This behavior is called ``down-sizing'' or
``anti-hierarchical'' evolution.  These results suggest that present-day
massive SMBHs formed in the early epoch of the Universe, compared with
less massive ones. A similar evolution is also observed in 
the star forming history of galaxies \citep[e.g.,][]{cowie1996, kodama2004}.
This provides a further evidence for the cosmological ``co-evolution''
scenario of SMBHs and galaxies, which is expected from the tight
correlation between a SMBH mass and galactic bulge properties observed
in the local Universe \citep[see e.g.,][]{magorrian1998, ferrarese2000, gebhardt2000}.

To elucidate the formation processes of SMBHs, it is very important to
reveal the AGN evolution at the high redshifts before the density peak, where
the rapid growth of SMBHs took place. Optical/infrared AGN surveys,
such as the Sloan Digital Sky Survey, the Canada-France High-z Quasar 
Survey, and the UKIDSS Large Area Survey, are able to detect very luminous 
quasars at high redshifts, now up to $z \simeq 7$ 
\citep{jiang2009, willott2010, mortlock2011}, revealing that their 
space number 
density has a peak around $z \simeq 2-3$ and shows a rapid decline 
by 96\% from $z = 3$ to $z = 6$ \citep{richards2006}. Recently, 
the space density of lower luminosity AGNs has been also investigated from deep
optical survey fields \citep{glikman2011, ikeda2011}, although the results are
still controversial due to the complexity of the corrections required for incompleteness and
contamination. While these optical surveys are useful in
detecting unobscured (so-called type-1) AGNs, they inevitably miss the
population of obscured (type-2) AGNs, the major class of the AGN
population \citep[see e.g.,][]{gilli2007}.

Hard X-ray observations at rest-frame energies above a few keV
provide an ideal opportunity to detect obscured AGNs, thanks
to their much smaller biases against absorption than the optical band and
low contamination from stars in the host galaxies. \cite{silverman2008} 
reports that the space density of X-ray-selected luminous AGNs
significantly declines at $z > 3$, similar to that of optically-selected
QSOs, although they have to assume a completeness correction 
because the spectroscopic identification rate of the sample 
is not high ($\sim50\%$). More recently,
\cite{brusa2009} and \cite{civano2011} have studied X-ray detected AGNs
at $z \simgt 3$ from the COSMOS survey \citep{scoville2007,
hasinger2007, elvis2009}, one of wide and deep 
multi-wavelength surveys over a continuous area. A decline of the space
density above $z \sim 3$ is confirmed for luminous AGNs with intrinsic
2--10 keV luminosities of $\simgt 10^{44}$ erg s$^{-1}$. 
While the functional shape of this decline is found to be
consistent with that of the exponential decline parameterization
proposed by \cite{schmidt1995}, which is based on the
optically-selected QSOs, it still has a  
large uncertainty due to the limited sample size.
For quantitative comparison with the results obtained in other wavelengths
and theoretical models, it is important to establish the shape of the decline
from a larger X-ray-selected AGN sample with high completeness.

Another key observational property of AGNs is the type-2 AGN
fraction (or more precisely, the distribution of absorption column
density), which gives strong constraints on the environments around
SMBHs including the obscuring tori and host galaxies. According to the
unified model of AGNs \citep{antonucci1993, urry1995}, the classification
of AGNs depends only on the viewing angle between the observer and the
rotation axis of the torus. It is well known that the type-2 AGN
fraction decreases with increasing X-ray luminosity \citep{ueda2003,
steffen2003, lafranca2005, hasinger2008, ueda2011}, indicating that the
simplest unified model with a fixed geometry of tori does not hold.
Several authors \citep[e.g.,][]{lafranca2005, treister2006,
hasinger2008} have furthermore reported that the type-2 fraction increases with
redshift, suggesting an evolution of the structures around SMBHs. 
However, due to
difficulties of type classification using photon-limited X-ray data and
complex biases against obscuration, the intrinsic type-2 AGN fraction at
$z>3$ has been only poorly investigated.

In this paper, we investigate the evolution of the comoving space
density of X-ray-selected luminous AGNs and type-2 fraction at high redshifts ($3
< z < 5$) using the X-ray catalog of Subaru/{\it
XMM-Newton} Deep Survey (SXDS), another wide and deep multi-wavelength
survey. The organization of the paper is as
follows. Section~\ref{sec:SAMPLE} describes the high-redshift AGN sample
used in this study. In section~\ref{sec:ANALYSIS}, we explain the
AGN type classification methods using X-ray or optical data. The
results of the comoving space density and type-2 fraction are presented
in section~\ref{sec:RESULT AND DISCUSSION}. Section~\ref{sec:SUMMARY}
summarizes the conclusions. Throughout this paper, the cosmological
parameters ($H_{\rm 0}$, $\Omega_{\rm M}$, $\Omega_{\Lambda}$) = (70 
km s$^{-1}$ Mpc$^{-1}$, 0.3, 0.7) are adopted. The ``log'' symbol
represents the base-10 logarithm. Quoted errors denote those at
$1\sigma$ level.

\section{SAMPLE}
\label{sec:SAMPLE}

\subsection{Identification of SXDS AGNs}

The Subaru/{\it XMM-Newton} Deep Survey \citep[SXDS;][]{furusawa2008,
ueda2008}
is one of wide and deep multi-wavelength survey projects covering the radio to
X-ray bands. The survey has an unprecedented combination of depth and
sky area over a contiguous region of $>$1 deg$^{2}$, centered at R.A. =
02$^{h}$18$^{m}$ and Dec. = $-05^{d}00^{m}$ (J2000). The main aims of
the SXDS survey are to explore the nature of extragalactic 
populations over the
whole history of the Universe, the co-evolution of galaxies and SMBHs, and the
evolution of the large-scale structure, without being affected by cosmic
variance.

{\it XMM-Newton} observed the SXDS field in seven pointings,
with one deep (nominal exposure of 100 ks) observation in the center 
surrounded by six shallower (50 ks each) ones,
in the 0.2--10 keV band with the European Photon Imaging Camera
\citep[EPIC;][]{struder2001, turner2001}. From the combined images
taken by three EPIC cameras (pn, MOS1, and MOS2) with
a total area of 1.14 deg$^{2}$,
\cite{ueda2008} detected 866, 1114, 645, and 136 X-ray sources with sensitivity 
limits of $6\times10^{-16}$, $8\times10^{-16}$, $3\times10^{-15}$, and
$5\times10^{-15}$ erg cm$^{-2}$ s$^{-1}$ in the 0.5--2, 0.5--4.5, 2--10,
and 4.5--10 keV bands, respectively, and with a detection likelihood $\ge$ 7
(corresponding to a confidence level of 99.91\%). 

Akiyama~et~al.\ (2012, in preparation) summarize the results of the identification 
of the SXDS X-ray sources detected in the 0.5--2 keV and/or 2--10 keV band
within the five combined fields-of-view of the Subaru/Suprime-Cam imaging, using
optical and infrared spectroscopic data as well as the deep imaging datasets
obtained with UKIRT/WFCAM and Spitzer/IRAC.
The 3$\sigma$ detection limit of the deep $R$-band image reaches 27.5 magnitude. 
Thanks to the deep optical image, objects with X-ray to optical flux ratio, 
log($f_{\rm X}/f_{\rm R}$), up to $+1.3$ can be detected, even for X-ray sources at the deepest 
X-ray detection limit in the soft X-ray band. The log($f_{\rm X}/f_{\rm R}$) distribution 
of typical AGNs \citep[$-1 < {\rm log}(f_{\rm X}/f_{\rm R}) < +1$;][]{akiyama2000} is well enclosed 
within the detection limit. Furthermore, the 3$\sigma$ limit of the $B$-band image 
(28.5 magnitude) is deep enough to detect the sharp decline below Ly$\alpha$ of
high-redshift objects with red $B-R$ color ($>$1.0 magnitude) 
even at the $R$-band detection limit.

Among the 866 sources detected in the 0.5--2 keV from
the whole SXDS field, 781 sources are
also covered with the deep Subaru/Suprime-Cam optical
images in $B$-, $V$-, $R$-, $i$-, and $z$-bands
\citep{furusawa2008}. In this study, we treat these
781 sources located within the overlapping region of the X-ray and 
deep optical imaging data as a parent sample.
Once candidates of Galactic stars, stellar objects that are close to a bright 
galaxy, and clusters of galaxies are removed, 733 out of the 781 sources 
are remained as candidates of AGNs. 
Among them, 586 are spectroscopically identified with intensive 
optical and near-infrared spectroscopic observations in the field. 
For the other AGN candidates without spectroscopic identification, 
their photometric redshifts are estimated on the basis of photometric data 
in 15 bands covering far-UV to mid-infrared wavelength range. 
Akiyama et al.\ (2012) apply HyperZ photometric redshift code \citep{bolzonella2000}, 
which determines a photometric redshift via template fitting with $\chi^2$ minimization, 
using both of galaxy and QSO Spectral Energy Distribution (SED) templates. 

In order to reduce the number of AGNs with photometric redshifts significantly different 
from spectroscopic ones (``outliers''), we apply two additional constraints in addition
to the $\chi^2$ minimization, considering the properties of the spectroscopically-identified 
AGNs. First one is that the objects with stellar morphology in the deep optical images 
are $z > 1$ broad-line AGNs. Almost all X-ray sources with stellar morphology 
are identified with type-1 AGNs at $z > 1$ in SXDS. They show bright nuclei and their 
observed optical lights are dominated by the nuclear components. This criterion may not be 
applicable to a survey with a deeper X-ray flux limit. Second one is the absolute magnitude 
range of the galaxy and QSO templates. Considering the absolute magnitude range of 
spectroscopically-identified type-1 and type-2 AGNs, we limit the $z$-band absolute magnitude 
ranges of the galaxy and QSO templates between $-25.0 < M_{z} < -20.0$ and $-26.5 < M_{z} < -22.0$, 
respectively. Further details of the photometric redshift determination are discussed in 
Akiyama et al.\ (2012).

%%%%%%%%%%%%%%%%
%   Figure 1   %
%%%%%%%%%%%%%%%%
%Fig.1
\begin{figure*}
  \epsscale{1.0}
  \plotone{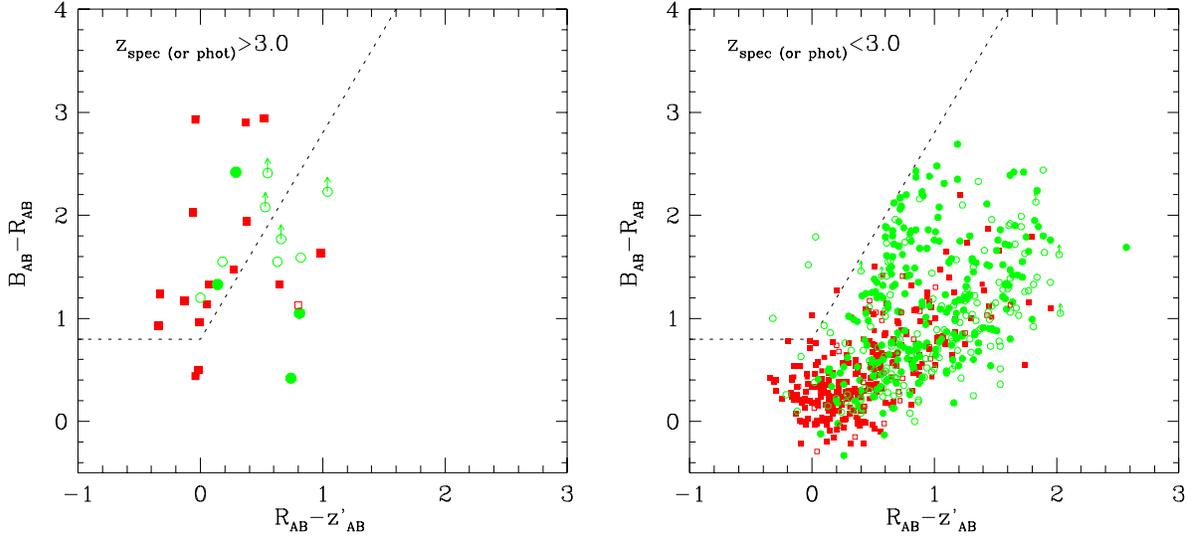}
  \caption{
    Distributions of SXDS AGNs on the $R - z$ and $B - R$
    color-color diagram (left: $z > 3$ objects, right: $z < 3$
    objects). The red squares and green circles represent type-1 and type-2
    AGNs, respectively. AGNs with spectroscopic and photometric
    redshifts are plotted with filled and open symbols, respectively.
    The arrows illustrate objects whose $B - R$ colors are their lower limits.
    The dashed line displays a possible criterion for $z > 3$ AGNs
    with dropout selection. \label{fig:SXDS_Bdr}}
\end{figure*}

The accuracy of the photometric redshifts is examined by comparing them with the spectroscopic 
ones. The median of $\frac{\Delta z}{(1+z_{\rm spec})}$ is 0.06 for the entire sample. 
We further examine the accuracy by the normalized median absolute deviation (NMAD; $\sigma_{z}$) 
following \cite{brammer2008}. For the entire sample, $\sigma_{z}$ is 0.104, 
which is larger than that of the photometric redshift estimations for X-ray-selected AGNs 
with medium band filters \citep{cardamone2010, luo2010}. 
The $\sigma_{z}$ for type-1 AGNs (0.201) is larger than that for type-2 AGNs (0.095).
This is because there is no strong feature in the SEDs of the type-1 AGNs except for the break 
below Ly$\alpha$. However, since the break can be covered by the deep optical images for AGNs 
at $z > 3$, the discrepancy between $z_{\rm spec}$ and $z_{\rm phot}$ is smaller for AGNs 
at $z > 3$. Regarding the completeness and contamination for the $z > 3$ AGN sample, 
one AGN at $z_{\rm spec} < 2.9$ shows $z_{\rm phot} > 3.0$. In contrast, five AGNs at 
$z_{\rm spec} < 2.9$ have $z_{\rm phot} > 3.0$.

It needs to be emphasized that in order to determine photometric redshifts of $z > 3$ 
AGNs reliably, deep optical imaging data are crucial since they can detect sharp decline 
of flux below the wavelength of rest-frame Ly$\alpha$ emission. 
Finally, most of the X-ray sources are identified as AGNs at
well estimated redshifts with a very high degree of completeness
($99.2\%$). Photometric redshifts cannot be determined for six 
objects that are detected in less than 4 bands. We discuss their
possible contributions in sections~\ref{sec:Pz} and \ref{subsec:space
density}.

Dropout method is a simple selection criterion for galaxies and AGNs at high redshifts, 
and sometimes such selection is preferred than the selection with the photometric redshifts, 
because the photometric redshifts can be affected by a catastrophic failure \citep{aird2010}. 
In order to compare the dropout selection with the photometric redshift selection, we plot 
$R-z$ and $B-R$ color-color diagram for AGNs at $z > 3$ (left) and $z < 3$ (right) 
in figure~\ref{fig:SXDS_Bdr}. The $z > 3$ AGNs show red $B-R$ color, which is consistent with 
the break below Ly$\alpha$ emission. The red color can be a crude check of the photometric 
redshifts, which are determined by the SEDs with all 15 bands. Most of the $z > 3$ type-1 AGNs 
have blue $R-z$ color as well, and they are enclosed within the criterion (dotted lines) that 
is determined to exclude most of the $z < 3$ AGNs (three $z < 3$ type-2 AGNs that are enclosed 
within the criterion have photometric redshifts of 2.6--2.8, close to $z = 3$). In contrast, 
$z > 3$ type-2 AGNs have redder $R-z$ color than type-1 AGNs on average, and some of them have 
similar colors to AGNs at lower redshifts. Thus, selection based on the optical colors only is 
useful in general but may miss a part of type-2 AGNs that have red UV color above Ly$\alpha$ 
emission.

\subsection{$z > 3$ AGN Sample}

For our study, we select a high-redshift AGN sample at $z > 3$ detected in
the 0.5--2 keV band, on the basis of spectroscopic and photometric redshifts
determined by Akiyama et al.\ (2012). Thanks to the K-correction effect,
even absorbed AGNs can be observed in the 0.5--2 keV band (corresponding
to the rest-frame energies above 2--8 keV for $z > 3$), where the
maximum sensitivity of {\it XMM-Newton} is achieved. 
Our final $z > 3$ AGN sample consists of 30
objects, among which 20 and 10 AGNs have spectroscopic and photometric
redshifts, respectively. The properties of the sample are summarized in
table~\ref{tab:sample} and the optical spectra of the spectroscopic 
sample are displayed in figure~\ref{fig:optical-spec}. 
Figure~\ref{fig:redshift_dist} exhibits the distribution of 
redshift for the $z > 3$ AGN sample.
Possible effects related to the uncertainties in 
the photometric redshifts are discussed in the next subsection.

%%%%%%%%%%%%%%%%
%   Figure 2   %
%%%%%%%%%%%%%%%%
%Fig.2
\begin{figure*}
  \begin{center}
    \epsscale{1.2}
    \plotone{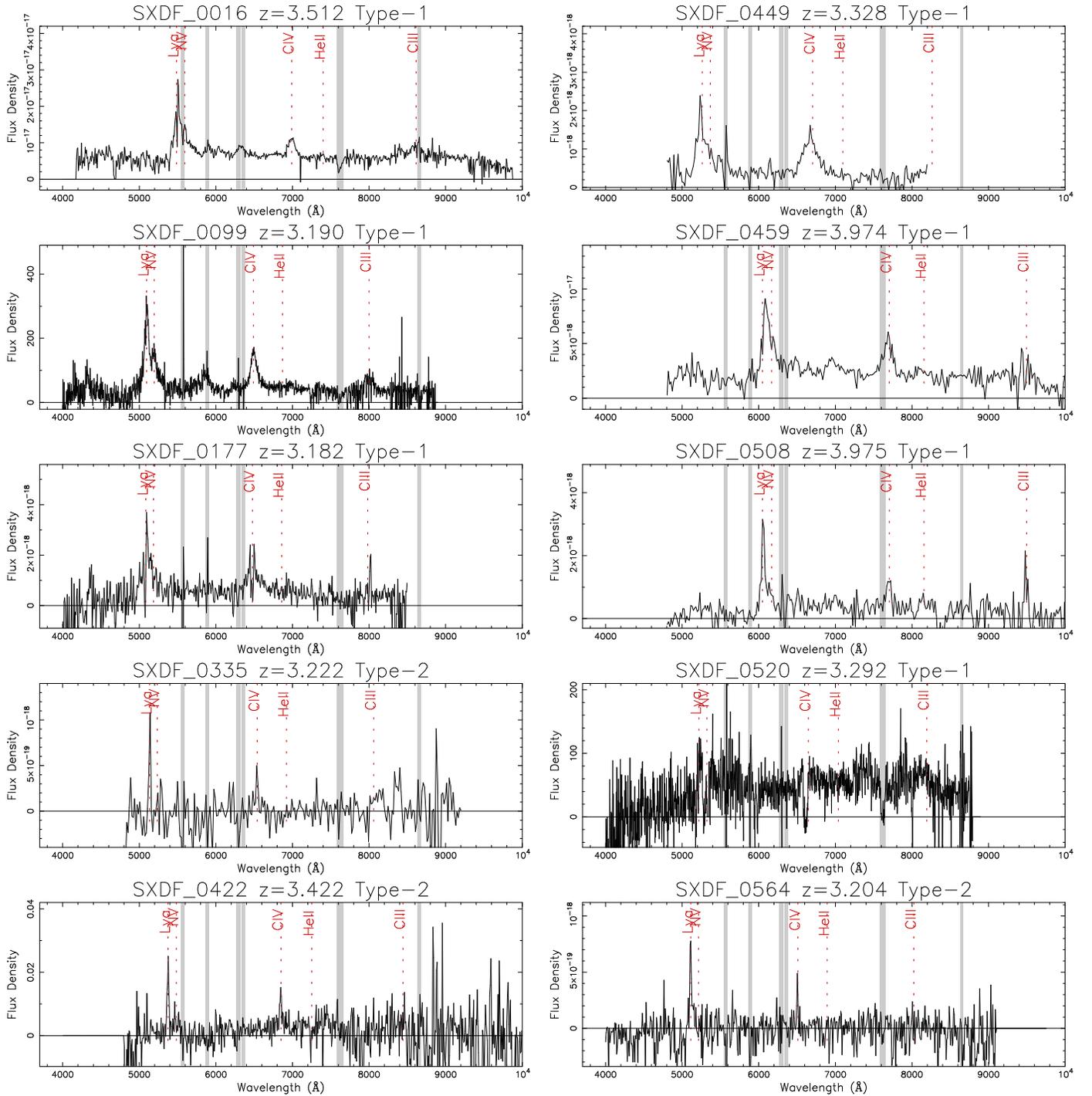}
    \caption{
      Optical spectra of the twenty $z > 3$ AGNs with spectroscopic
      redshifts. The vertical axis represents the flux density per
      wavelength (some spectra are not flux-calibrated). The vertical dashed
      lines in each panel indicate the positions of emission lines typically
      seen in AGNs. The gray hatched regions exhibit wavelength ranges
      affected by strong night sky lines or atmospheric absorption
      lines. \label{fig:optical-spec}}
  \end{center}
\end{figure*}

\addtocounter{figure}{-1}
\begin{figure*}
  \begin{center}
    \epsscale{1.2}
    \plotone{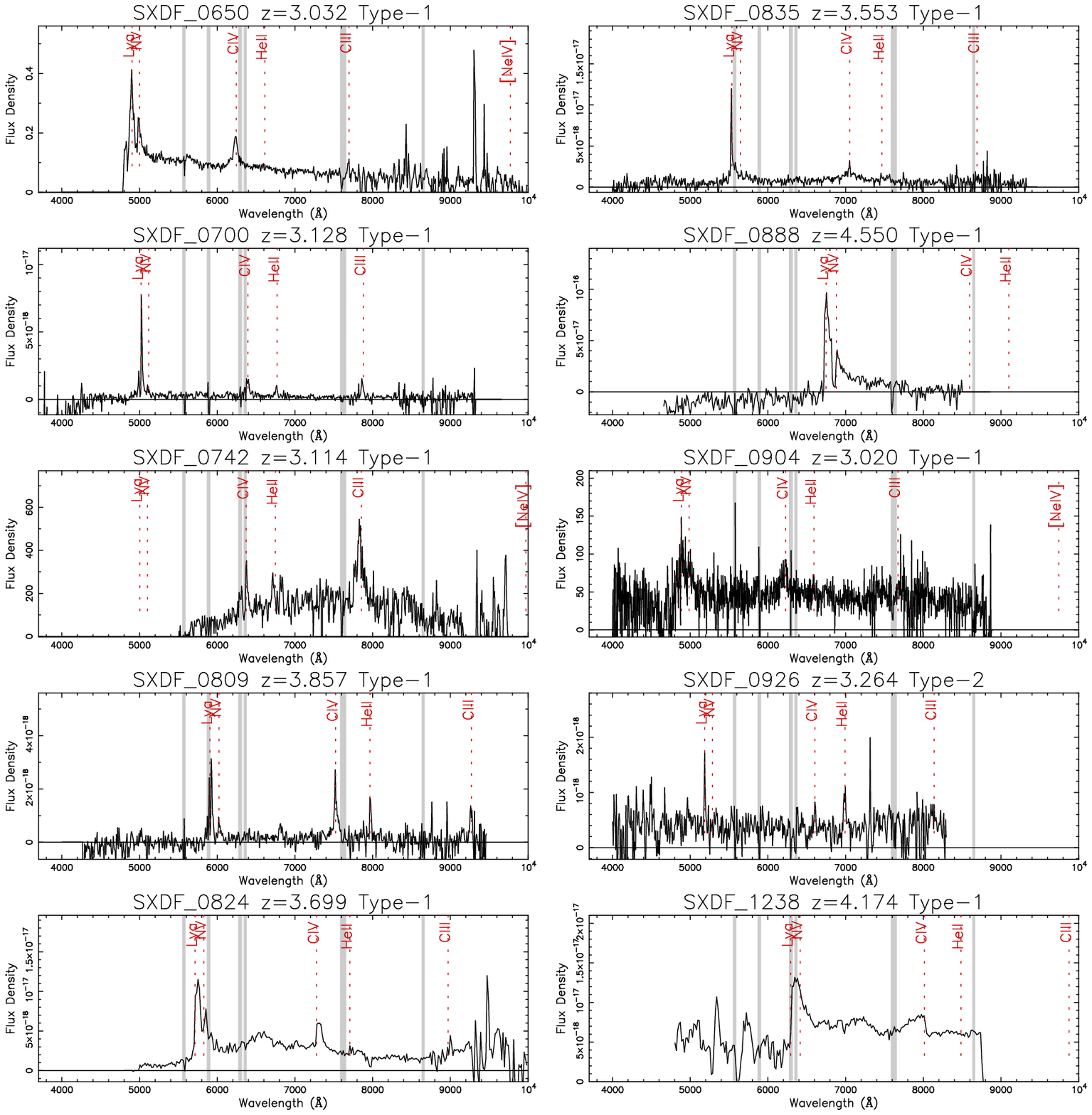}
    \caption{{\it Continued.}}
  \end{center}
\end{figure*}

%%%%%%%%%%%%%%%%
%   Figure 3   %
%%%%%%%%%%%%%%%%
%Fig.3
\begin{figure}
  \epsscale{1.2}
  \plotone{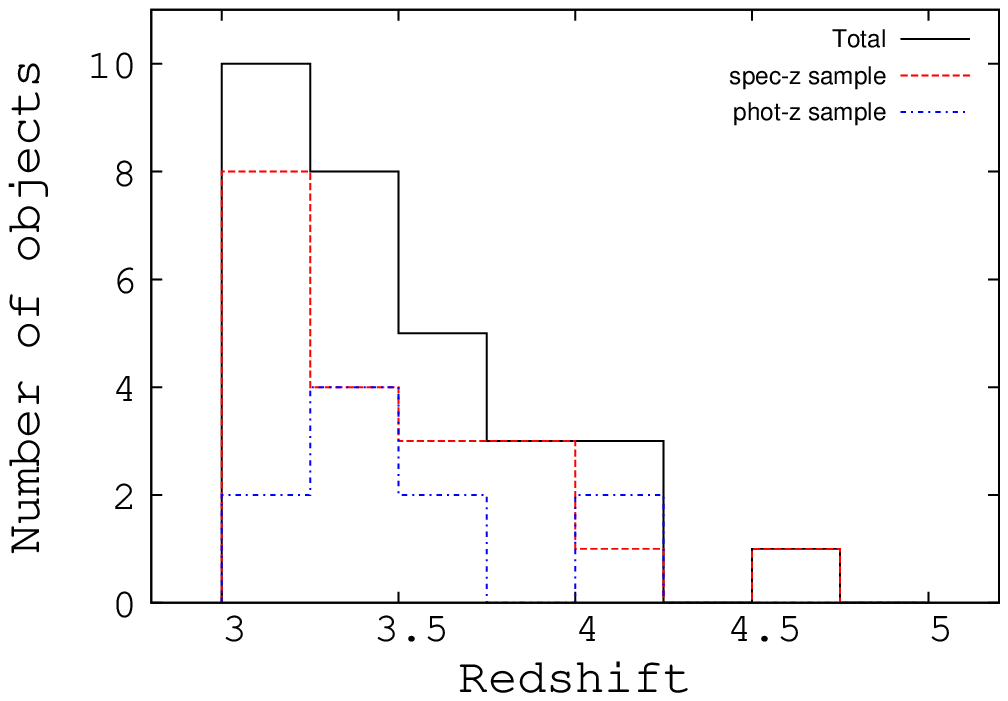}
  \caption{Redshift distribution for $z > 3$ AGN sample. Black solid, red dashed, and
    blue dot-dashed lines represent the total, spectroscopic redshift, and photometric redshift
    samples, respectively. \label{fig:redshift_dist}}
\end{figure}

\subsection{Uncertainties of Photometric Redshifts}
\label{sec:Pz}

%%%%%%%%%%%%%%%%
%   Figure 4   %
%%%%%%%%%%%%%%%%
%Fig.4
\begin{figure}
  \epsscale{2.0}
  \plottwo{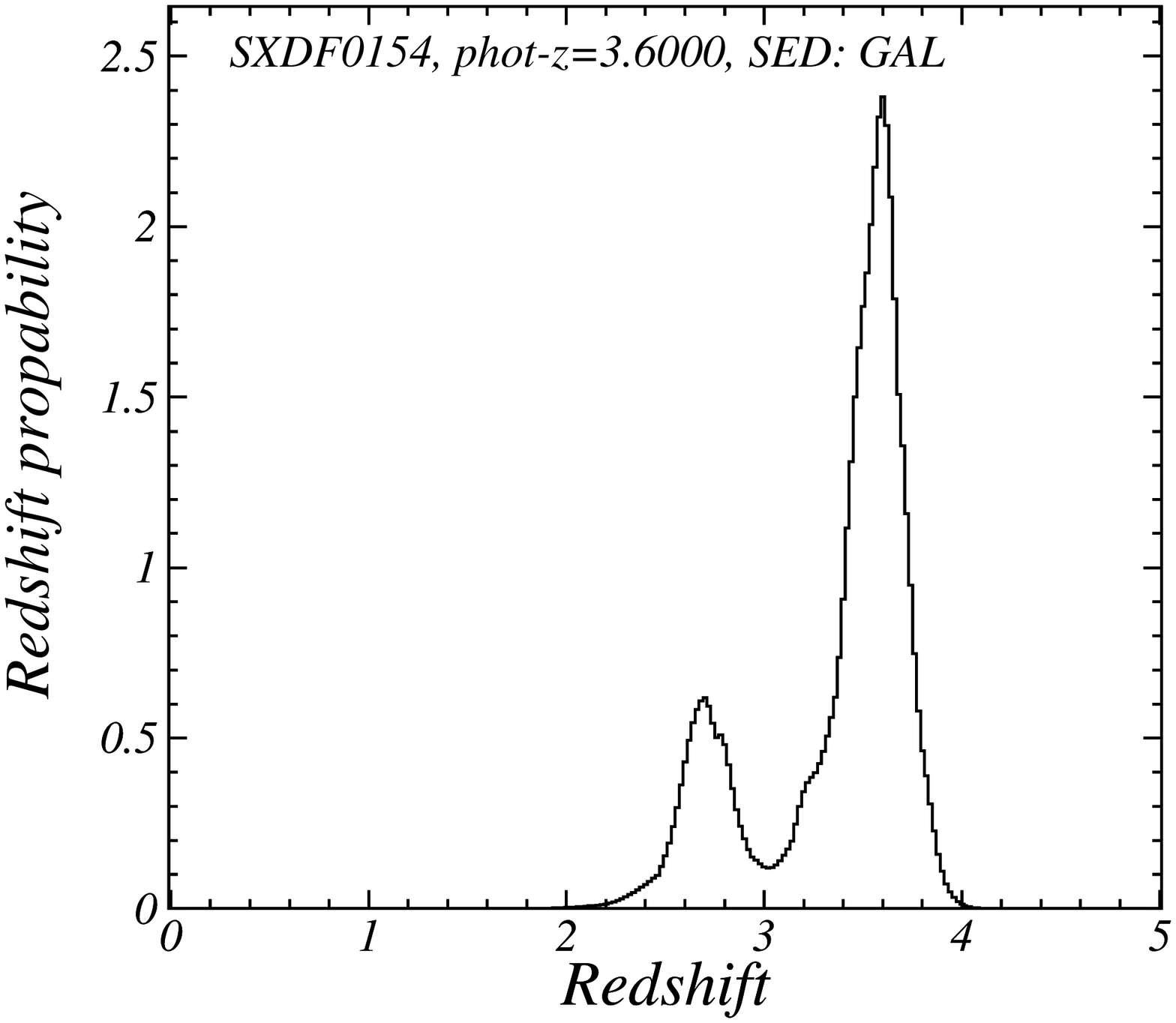}{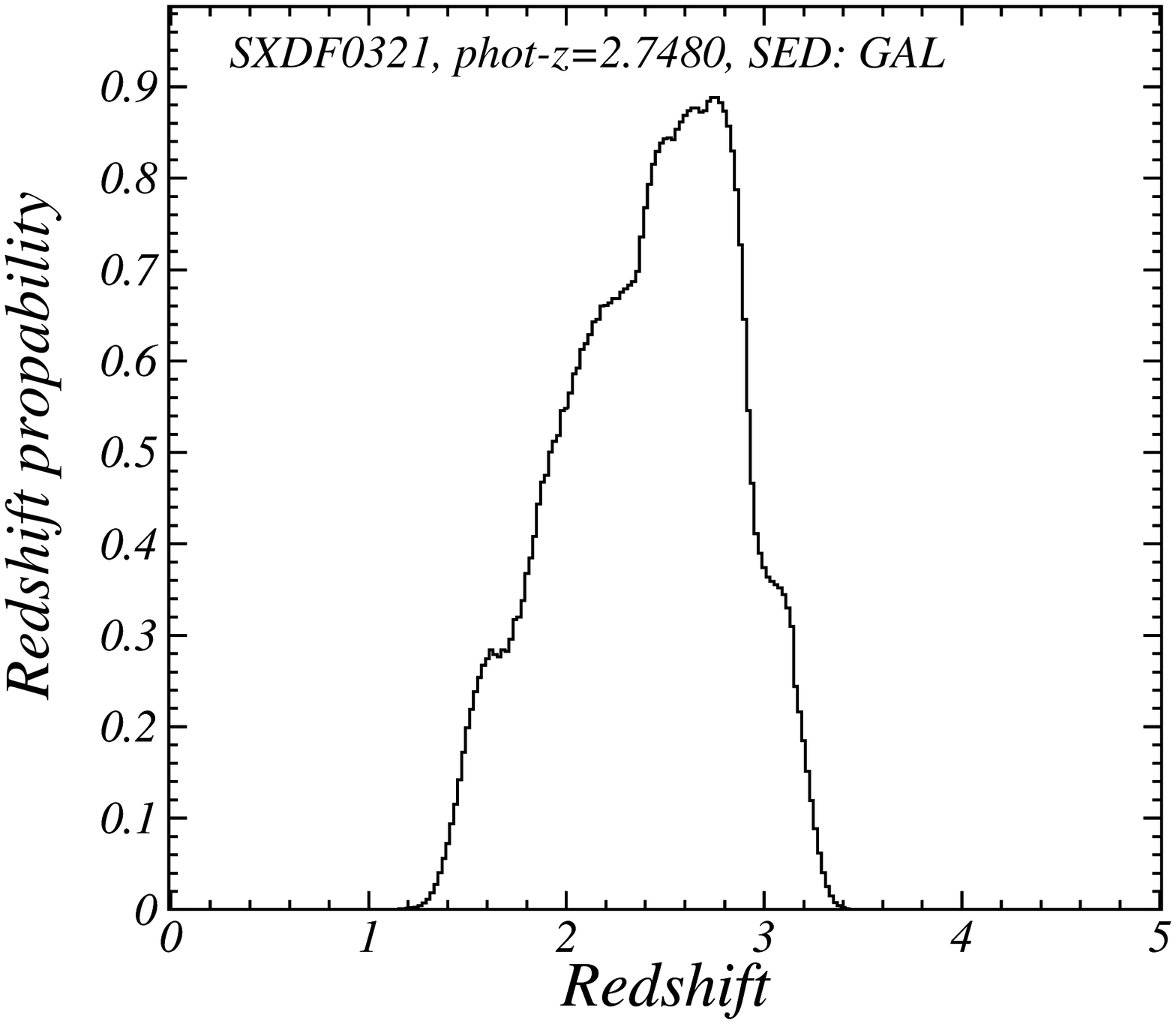}
  \caption{Examples of $P(z)$, the probability distribution function for
    photometric redshift solution (top: SXDF0154, bottom:
    SXDF0321). \label{fig:PDFs} Distributions are normalized so that $\int_{z =
    0}^{5} P(z)dz = 1$.  }
\end{figure}

Since the photometric redshifts are determined only by
broad-band photometries, they could have uncertainties unlike
spectroscopic redshifts.
Figure~\ref{fig:PDFs} illustrates examples of the normalized 
probability distribution functions (PDFs, P($z$)) 
for the photometric redshift for two sources without spectroscopic redshifts, 
SXDF0154 and SXDF0321, calculated from the distribution of the $\chi^2$ value
obtained in the SED fitting. SXDF0154 has a primary solution (i.e.,
the highest peak of PDF) at $z > 3$, and hence is included in our $z > 3$ 
AGN sample. By contrast, SXDF0321 is not included in the sample since
the PDF has its peak at $z < 3$. Both of them, however, have finite
probabilities at $z < 3$ and $z > 3$.

To take into account the shape of the PDF in each object statistically,
we consider the contribution from all 0.5--2 keV selected AGNs in the 
Akiyama et al.\ (2012) catalog to the $z > 3$ AGN sample,
using the normalized PDFs between $z = 0$ and $z = 5$.
For the sources with spectroscopic redshifts, we define their PDFs as
delta-functions peaked at their measured redshifts. It is found that the
``effective'' number of objects in the $z > 3$ AGN sample, computed by
summing the fraction of the normalized PDFs in the $z$ = 3--5 region
($\int_{3}^{5} P(z) d z$) from all objects, is
30.6, when the six unidentified objects are ignored. This value is
almost the same as the number of the best-fit $z > 3$ AGN sample
described in the previous subsection. Even if we include the
unidentified objects for the calculation, the effective number is
computed to be 33.7, which is still within the 1$\sigma$ error of the
number of our $z > 3$ AGN sample. We also confirm that there are no
significant changes in the luminosity and absorption
column density distributions of the sample when we consider the PDFs of
photometric redshifts; here the contribution of each
object to the distributions is added with a weight of $P(z)$ over the
redshift range of $z$ = 3--5. Hence, we hereafter consider only the
best-fit values of the photometric redshifts.

%%%%%%%%%%%%%%%
%   Table 1   %
%%%%%%%%%%%%%%%

\section{ANALYSIS}
\label{sec:ANALYSIS}

\subsection{Classification of AGNs}
\label{sec:Classification of AGNs}

For our discussion of the obscured AGN fraction, we need to classify each
object in our sample as a type-1 or type-2 AGN. There are two
different ways to classify AGN: one is based on the optical spectra (optical
type) and the other is based on the absorption column densities, $N_{\rm H}$, in the X-ray
spectra (X-ray type). The optical type is basically determined by the
presence of broad, permitted emission lines. 
Considering the distribution of the full width at half maximum (FWHM) 
of the broad-line AGNs in the local universe \citep{stern2012}, we classify our spectroscopic 
sample: those showing significant broad lines with velocity widths larger than 1,000 km s$^{-1}$ 
are optical type-1 and the others are optical type-2. In the $z > 3$ sample, all but one 
(SXDF0809) broad-line AGNs have the FWHM of the broad-line width larger than 2,000 km 
s$^{-1}$. Thus, their optical classifications are not changed even if we use the threshold 
of 2,000 km s$^{-1}$ instead of 1,000 km s$^{-1}$. The CIV line of SXDF0809 has the FWHM of 
1,470 km s$^{-1}$. Although it is classified as a type-1 AGN since the CIV line shows broader wing, 
it needs to be noted that it also shows a strong HeII line, which is typical for narrow-line AGNs. 
The contribution from the narrow-line component is large in this object. 

For the 10 AGNs without spectroscopic redshifts
(i.e., those that do not show any significant line features or do not
have optical/near-infrared spectra available), we classify them by the
type of a template spectrum (i.e., QSO or galaxy).  For
spectroscopically-identified AGNs, there is a good correlation between
the spectral type and the best-fit SED type: type-1 (type-2) AGNs are
well fitted with QSO (galaxy) templates.  Specifically, among the
twenty $z > 3$ AGNs with spectroscopic identification, the
spectroscopic and photometric types match well for 15 objects.  We
therefore classify spectroscopically-unidentified objects fitted
better with QSO (galaxy) template into type-1 (type-2) AGNs.
Considering the mismatch rate of 25\%, we estimate
that less than two (= 9*0.25 - 1*0.25) out of the nine AGNs with galaxy spectra could be
truly type-1 AGNs; this uncertainty will be taken into account in the
discussion on optical type-2 AGN fraction (section~\ref{sec:opt_f_type2}). 
We do not consider combined templates with galaxy and AGN in the photometric
redshift determination. It will not affect the classification of the
$z > 3$ AGNs, since at $z > 3$ the SXDS sample covers only luminous
AGNs and the SEDs of luminous type-1 AGNs are completely dominated by
QSO components.

A general problem of relying on the 
optical type is that the classification may depend on the quality of
the available optical spectra, since good signal-to-noise ratio is
required to detect less-luminous broad-emission lines on top of stellar 
continuum emission of host galaxies. Nevertheless, such effects are not
expected to be significant in our sample since it consists of
luminous AGNs for which contamination from the host galaxies is
negligible.

X-ray absorption is an alternative good indicator of AGN type, 
as expected from the unified scheme \citep{antonucci1993, urry1995}
: type-2 AGNs are viewed through the dust torus, and hence have higher 
absorption column densities than type-1 AGNs. In the present work, we
adopt $N_{\rm H} = 10 ^{22}$ cm$^{-2}$ as the dividing criterion: AGNs
with log $N_{\rm H} < 22$ and $> 22$ are classified as X-ray type-1 and
type-2, respectively. This criterion is adopted by many authors, and is
known to be generally in good agreement with the optical type 
\citep[see e.g.,][]{ueda2003}. The merit of using 
X-ray classification is that observational biases in X-ray surveys
can be well defined and can be corrected in a straightforward way.
Thus, we basically adopt the X-ray type for our study, although 
consistency with results based on the optical type is also discussed.

\subsection{Absorption Column Density}
\label{sec:Absorption Column Density}

\begin{figure}
  \epsscale{1.2}
  \plotone{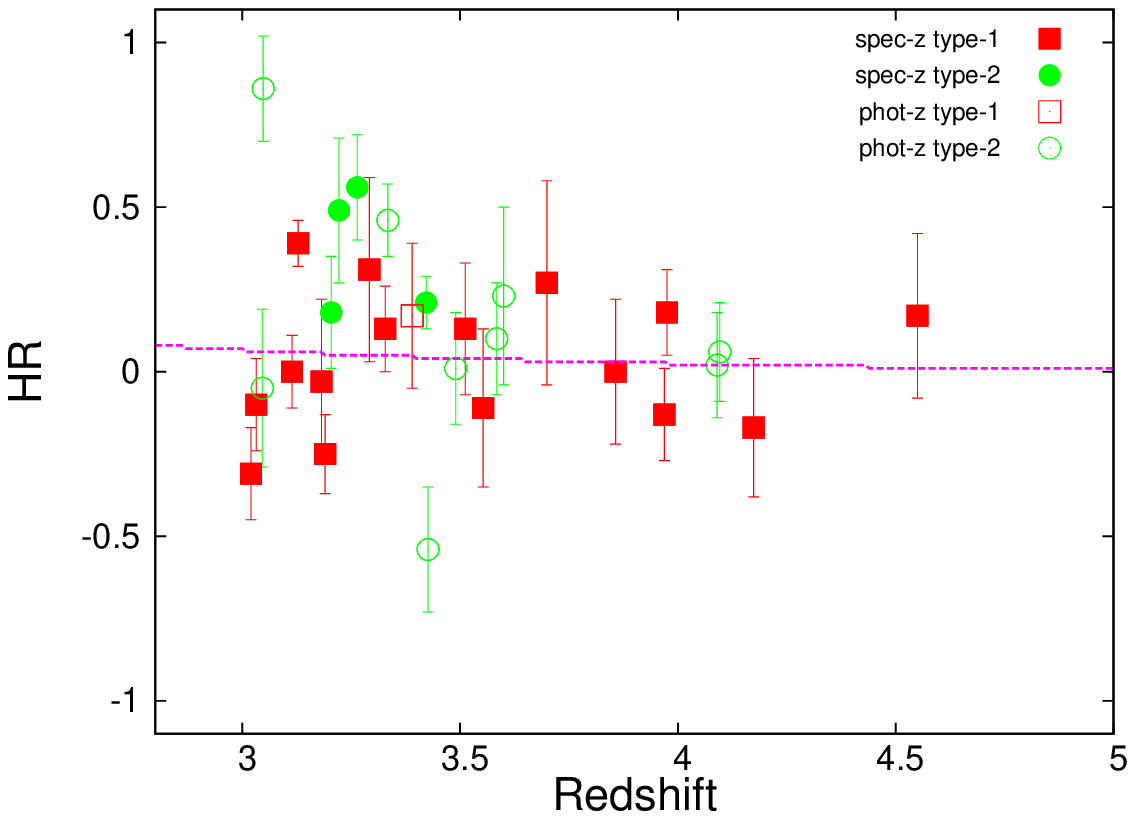}
  \caption{Hardness ratio (HR) versus redshift plot for our AGN sample.
    The red and green symbols correspond to type-1 and type-2
    AGNs, respectively. The filled and open circles represent spectroscopic and
    photometric redshifts samples, respectively.
    The magenta dashed line corresponds to log \nh\ = 22.
    \label{fig:HR4-z_opttype}}
\end{figure}

To derive the absorption column density ($N_{\rm H}$) for each AGN, we
use an X-ray hardness ratio (HR) between two energy bands, since it
is difficult to perform spectral fitting due to the limited photon
statistics for most of the X-ray sources in our sample.
We define the HR value as
\begin{equation}
  {\rm HR} = (H-S)/(H+S),
\end{equation}
where $S$ and $H$ are the vignetting-corrected count rates
\footnote{corresponding to those of the pn detector, obtained by
dividing the combined pn+MOS1+MOS2 image by the pn-equivalent exposure map;
for details see \cite{ueda2008}} in the 0.3--1.0 keV and 1.0--4.5 keV
bands, respectively. We find that this choice of bands gives the best
sensitivity for the determination of the $N_{\rm H}$ value 
for $N_{\rm H}\sim 10^{22}$ cm$^{-2}$,
for objects at $z$ = 3--5. Since these count rates are not available in the 
\cite{ueda2008} catalog, we re-analyzed the {\it XMM-Newton} images in
these two bands at the fixed source positions, utilizing exactly the same
datasets and analysis software as used in \cite{ueda2008}. The relationship
between HR and redshift of our AGN sample is shown in
figure~\ref{fig:HR4-z_opttype}.

%%%%%%%%%%%%%%%%
%   Figure 5   %
%%%%%%%%%%%%%%%%
%Fig.5

%%%%%%%%%%%%%%%%
%   Figure 6   %
%%%%%%%%%%%%%%%%
%Fig.6
\begin{figure}
  \epsscale{1.2}
  \plotone{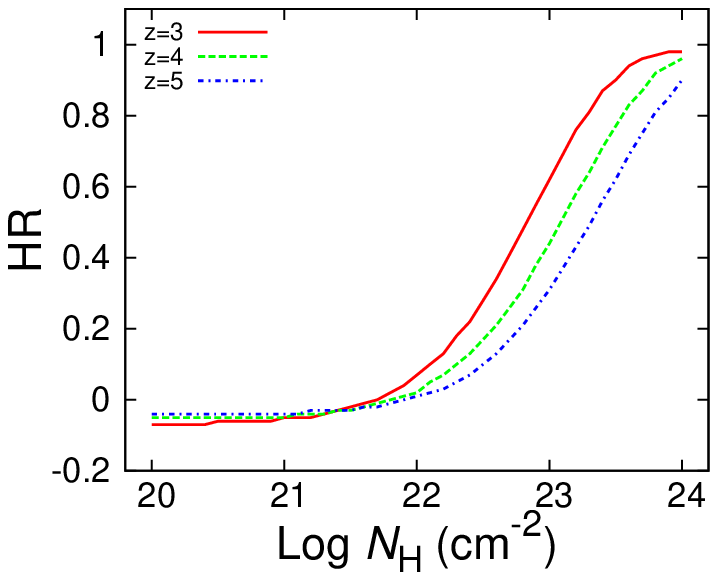}
  \caption{The relationship between the hardness ratio HR and absorption
  column density $N_{\rm H}$.  A power law with a photon index of 1.9
  plus a reflection component from cold matter with the solid angle
  $\Omega=2\pi$ is assumed as the intrinsic spectrum. The red solid, green
  dashed, and blue dot-dashed lines correspond to the redshifts of 3, 4,
  and 5, respectively. \label{fig:HR4-logNH_relation}}
\end{figure}

\begin{figure}
  \epsscale{1.2}
  \plotone{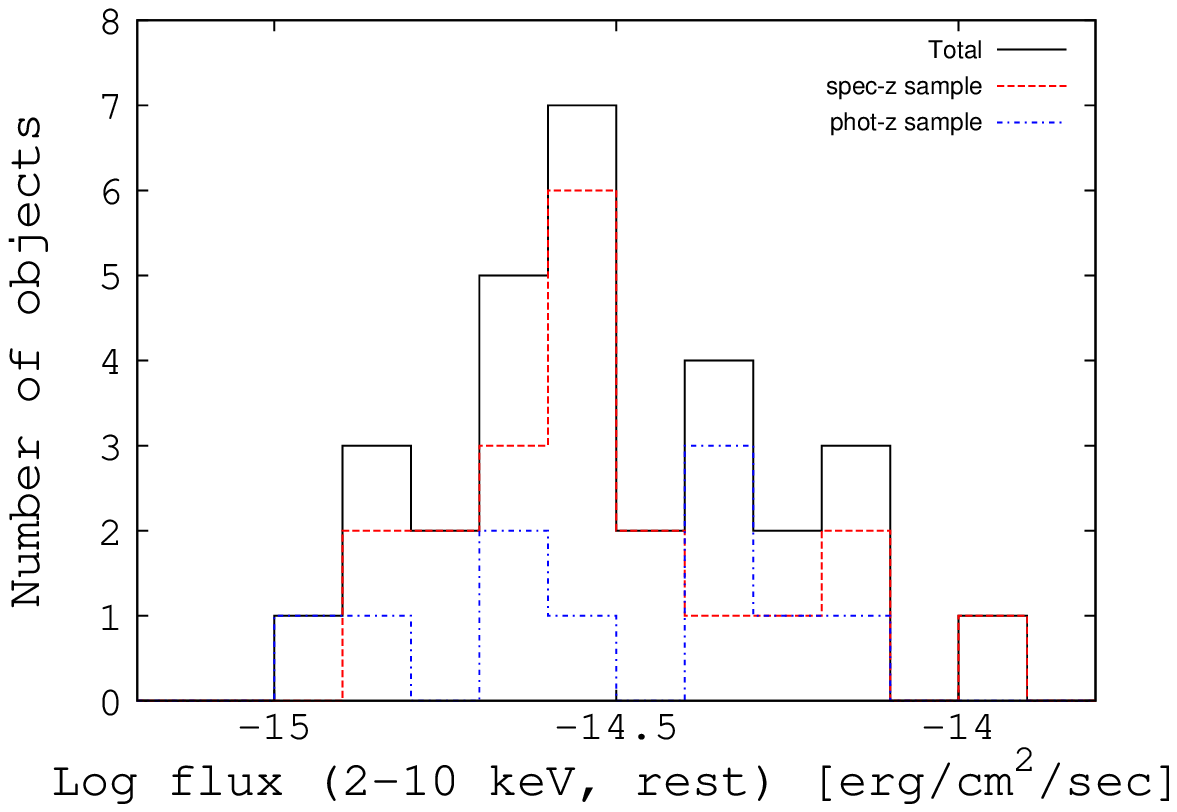}
  \caption{Flux distribution for $z > 3$ AGN sample.
      The definition of the symbols is the same as in figure~\ref{fig:redshift_dist}.
    \label{fig:flux_dist}}
\end{figure}

\begin{figure}
  \epsscale{1.2}
  \plotone{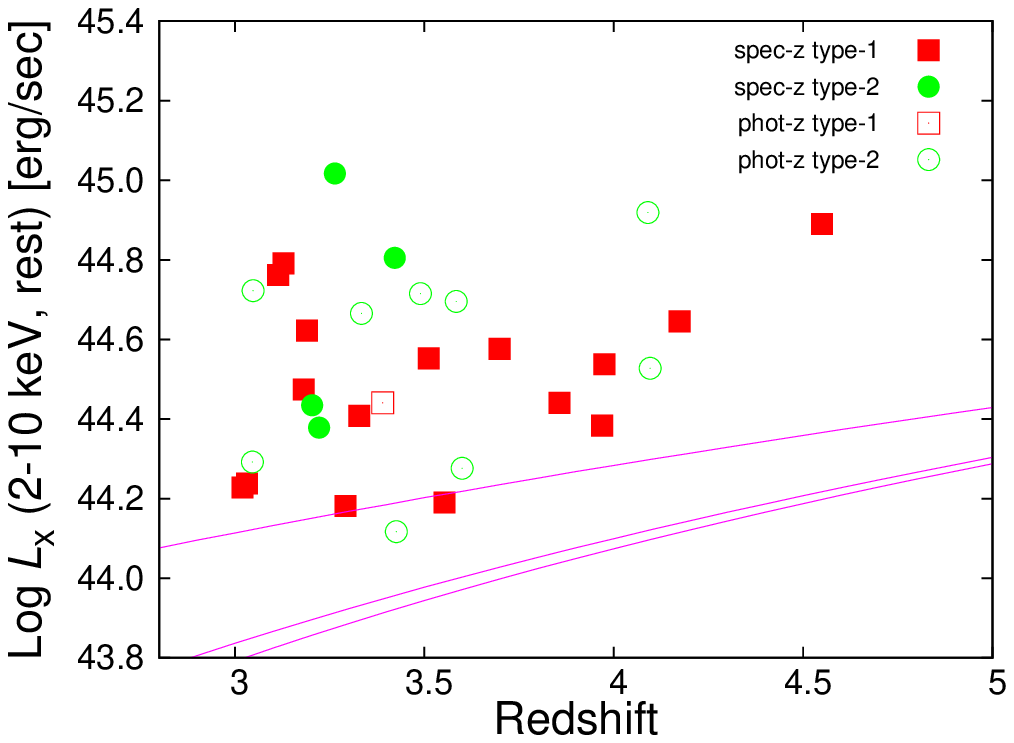}
  \caption{Correlation between the intrinsic (de-absorbed and rest-frame
   2--10 keV) luminosity and redshift for our AGN sample. The definition of the
   symbols is the same as in figure~\ref{fig:HR4-z_opttype}. The magenta
   lines represent the detection limits for AGNs with log $N_{\rm H}$ =
   20, 22, and 23, from bottom to top, calculated from the 0.5--2 keV
   count rate corresponding to one tenth of the total survey area. \label{fig:logLx-z_opttype}}
\end{figure}

Figure~\ref{fig:HR4-logNH_relation} shows the relationship between HR and
column density at $z$ = 3, 4, and 5. Here and after, 
we assume a power law with a 
photon index of 1.9 plus a reflection component from cold matter with a
solid angle of $2\pi$ as the intrinsic spectrum, following \cite{ueda2003}.
We derive the column densities of our AGN sample from the redshift and
observed HR value using this relation, and then calculate their
intrinsic (de-absorbed and rest-frame 2--10 keV) fluxes and luminosities. 
Where an observed HR corresponds to log $N_{\rm H} < 20$, we set it
to log $N_{\rm H} = 20$. The column density, flux, and luminosity are
also summarized in table~\ref{tab:sample}, and the flux 
distribution is presented in figure~\ref{fig:flux_dist}.
Figure~\ref{fig:logLx-z_opttype} displays the relationship 
between the intrinsic luminosity and redshift for our AGN sample. The magenta 
curves represent the detection limits for AGNs with log $N_{\rm H}$ = 
20, 22, and 23, calculated from the 0.5--2 keV count rate 
at which the survey area becomes one tenth of the maximum value. 
As noticed from the figure, the survey is incomplete for sources with log $N_{\rm H} > 23$ 
in the lowest luminosity range below log $L_{\rm X} \sim 44.2$ at $z > 3.5$.
However, we confirm that this does not affect our estimate on 
the space density of all Compton-thin AGNs with log $N_{\rm H} < 24$ 
over the statistical errors as far as the fraction of heavily obscured AGNs 
with $N_{\rm H} = 23-24$ is not extremely high.
We find that all of our AGN sample have intrinsic luminosities of log
$L_{\rm X} > 44$ erg s$^{-1}$, and hence belong to the QSO class.
While the X-ray AGN types match the optical ones for 18 out of 30 AGNs 
in our sample, they do not for the others. This is largely due to 
the statistical error of column density. We can, however, correct for it 
by considering a conversion matrix from an intrinsic to observed 
$N_{\rm H}$ distributions (see section~\ref{subsub:method2}).

%%%%%%%%%%%%%%%%
%   Figure 7   %
%%%%%%%%%%%%%%%%
%Fig.7

%%%%%%%%%%%%%%%%
%   Figure 8   %
%%%%%%%%%%%%%%%%
%Fig.8

\section{RESULT AND DISCUSSION}
\label{sec:RESULT AND DISCUSSION}

\subsection{Evolution of the Comoving Space Density of AGNs at $z > 3$}
\label{subsec:space density}

To investigate the cosmological evolution of AGNs at $z>3$, we calculate
the comoving space density from our sample utilizing the $1/V_{\rm max}$
method \citep{schmidt1968}. This method takes into account the fact
that a fainter source can be found in a smaller volume because of the 
survey area dependence on flux (or more correctly, count rate).
The maximum available volume $V_{\rm max}$ is calculated by 
the following formula:
\begin{equation}
  V_{\rm max} = \int_{z_{\rm min}}^{z_{\rm max}} \Omega(c(L_{\rm X}, z, N_{\rm H})) (1 + z)^{3} d_{A}(z)^{2} c\frac{d\tau}{dz}(z) dz,
\end{equation}
where $\Omega(c(L_{\rm X}, z, N_{\rm H}))$ is the sky coverage at 
the count rate $c(L_{\rm X}, z, N_{\rm H})$ expected 
from a source at the redshift $z$ with the intrinsic
luminosity $L_{\rm X}$ and the absorption column density $N_{\rm H}$,
$d_{A}$ and $c\frac{d\tau}{dz}$ are the angular distance and the
look-back time distance per unit redshift, respectively, both of which
are functions of $z$. Here $z_{\rm min}$ and $z_{\rm max}$
are the lower and the upper edges of the redshift bin used for computing
$V_{\rm max}$, respectively. In the present paper, we separate the
redshift interval of $3 < z < 5$ into five bins ($z$ = 3.0--3.2,
3.2--3.4, 3.4--3.8, 3.8--4.3, and 4.3--5.0),
and calculate $V_{\rm max}$ for each AGN 
in the corresponding redshift bin.
After calculating $V_{\rm max}$, we sum up the reciprocal value in each redshift bin:
\begin{equation}
  \phi = \sum_{i}^{z_{\rm min} < z < z_{\rm max}} \left(\frac{1}{V_{{\rm max}, i}}\right),
\end{equation}
where $\phi$ is the comoving space density in the range $z$ = $z_{\rm
min}$--$z_{\rm max}$, and $i$ is the index of the sample AGN falling into
the redshift bin. The 1$\sigma$ error is estimated as:
\begin{equation}
  \sigma\phi = \sqrt{\sum_{i}^{z_{\rm min} < z < z_{\rm max}} \left(\frac{1}{V_{{\rm max}, i}}\right)^{2}}.
\end{equation}
Where only one source occurs in a given redshift bin, we adopt the
Poisson upper and lower limits corresponding to the confidence range of
Gaussian 1$\sigma$ formulated by \cite{gehrels1986}.

%%%%%%%%%%%%%%%%
%   Figure 9   %
%%%%%%%%%%%%%%%%
%Fig.9
\begin{figure}
  \epsscale{1.2}
  \plotone{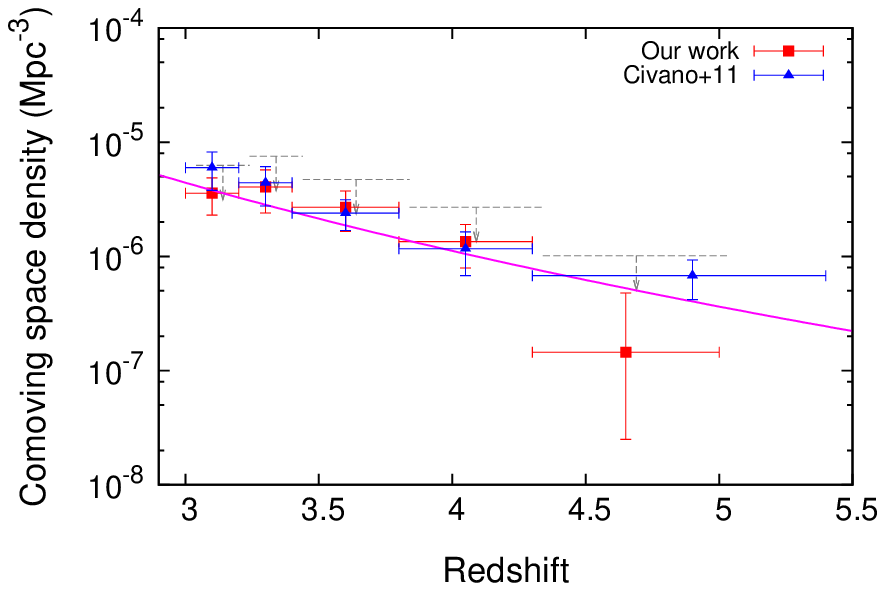}
 \caption{Comoving space density of X-ray-selected, luminous AGNs
 plotted as a function of redshift.  The red squares and blue triangles
 indicate our results at log $L_{\rm X}$ $\ge$ 44 and those of
 \cite{civano2011} at log $L_{\rm X}$ $\ge$ 44.15, respectively. The
 errors shown are 1$\sigma$. The gray dashed arrows represent the strict
 upper limits at each redshift bin (for visuality, they are slightly shifted
 in the x-axis direction),  calculated by assuming that all of the unidentified 6 objects are included
 in the bin. The magenta curve corresponds to the
 best-fit model fitted to the SXDS data with the form of $(1 +
 z)^{-6.2}$, obtained from the simultaneous fit. \label{fig:compare_zoom}}
\end{figure}

Figure~\ref{fig:compare_zoom} displays, as a function of redshift, the
comoving space density of AGNs with luminosities log $L_{\rm X}$ =
44--45 erg s$^{-1}$ obtained from the SXDS sample (red squares).
The gray dashed arrows represent the strict upper
limits at each redshift bin, calculated by assuming that all of the
unidentified 6 objects are included in the bin. The results from the
COSMOS survey \citep{civano2011} for AGNs with log $L_{\rm X} \ge
44.15$ erg s$^{-1}$ are also plotted for comparison (blue triangles).
As can be seen, our results are consistent with those of COSMOS within
the statistical errors, indicating a significant decline in the AGN
space density from $z = 3$ to higher redshifts. 

When we fit the SXDS result with a power-law form of $\propto (1 +
z)^{p3}$, we obtain $p3 = -7.0 \pm 1.8$.
We confirm that the index does not change
over the error even when we consider the PDFs 
of photometric redshifts for all objects including the six 
unidentified ones (see section~\ref{sec:Pz}).
A simultaneous fit to the SXDS and COSMOS data yields a best-fit slope of 
$p3 = -6.2 \pm 0.9$, which is shown by the solid magenta curve in
figure~\ref{fig:compare_zoom}. 
We find that an exponentially
decaying model above a critical redshift $z_{c2} = 2.7$, with the form
$10^{p4(z-z_{c2})}$ as adopted by \cite{schmidt1995}, also describes
these data well; we obtain $p4 = -0.65 \pm 0.15$ (SXDS only) and
$p4 = -0.60 \pm 0.08$ (SXDS+COSMOS). These results on $p3$ or $p4$ give
the tightest constraints on the decay profile of the space density of
X-ray-selected luminous AGNs with log $L_{\rm X}$ = 44--45 erg s$^{-1}$ by
utilizing a total of 80 objects.
The parameters are little affected by filtering the SXDS sample with
log $L_{\rm X}$ $\ge$ 44.15, the same criterion for the COSMOS sample.

We confirm that the shape of the space-density evolution of X-ray-selected 
luminous AGNs is consistent with that derived from optical QSO
surveys within current uncertainties. With the exponential model,
\cite{schmidt1995} obtained $p4 = -0.43\pm0.04$ from optically-selected QSOs
with $M_B < -26$, which roughly corresponds to log $L_{\rm X} > 44.5$
assuming a typical SED of QSOs \citep[see e.g.,][]{ueda2003}. More recent
results by \cite{richards2006} give $p4 \simeq -0.67$ for optical QSOs at a
magnitude limit of $M_{i} < -27.6 $. Our result ($p4 = -0.60 \pm 0.08$)
obtained from X-ray surveys for AGNs with log $L_{\rm X} > 44$ lies between 
the two optical results, as already noted by 
\cite{brusa2009}. Since X-ray-selected samples contain both type-1 and
type-2 AGNs, the same decline
profile as that of optically-selected
(type-1) QSOs suggests no significant cosmological evolution of the
type-2 AGN fraction above $z > 3$, at least in the bright luminosity range. 
This result is consistent with the conclusion by \cite{hasinger2008} 
that the type-2 AGN fraction increases with $z$ but saturates beyond $z=2$ or
$z=3$.

\subsection{Fraction of Type-2 AGNs at $z$ = 3--5}

To investigate the cosmological evolution 
of the fraction of type-2 AGNs ($f_{\rm type-2}$) in the whole Compton-thin population 
(i.e., with $N_{\rm H} < 10^{24}$ cm$^{-2}$) from $z=0$ to $z=5$, 
we constrain $f_{\rm type-2}$ at $z$ = 3--5 
in the intrinsic luminosity range of log $L_{\rm X}$ = 44--45 using our 
SXDS sample.
We assume no redshift evolution within the redshift 
range of $z$ = 3--5, following the previous discussion. There
are observational biases in the observed ratio of type-1 and
type-2 AGNs, which require careful correction to obtain the true
fraction. As mentioned in section~\ref{sec:Classification of AGNs}, 
we primarily adopt the ``X-ray
type'' defined by the absorption seen in the X-ray spectra, for which 
an evaluation of such biases can be made without assuming a relationship
between the optical and X-ray properties. We finally discuss the
cosmological evolution of the type-2 AGN fraction by comparing our
work with previous results obtained at lower redshifts.

\subsubsection{X-ray Type-2 Fraction}
\label{subsub:method2}

In our analysis, we consider the two effects. (1) For a given
intrinsic luminosity, X-ray type-2 AGNs become less detectable due to
the reduction of the observed count rate by photoelectric
absorption. (2) Due to the limited number of photons detected, there
is a statistical fluctuation of the best-fit column density derived
from the observed hardness ratio (i.e., an AGN can apparently have a
different absorption column density from the true
value). It must be noted that the observed
column densities are not simply subject to Gaussian or Poissonian 
errors but have complex probability distributions 
with unsymmetric negative and positive errors.
This could make the ``observed'' distribution 
simply based on the best-fit column density of each object
quite different from the ``intrinsic'' one.

To obtain the true absorption distribution by taking
into account these two biases simultaneously, we consider a
``conversion matrix'' from the intrinsic $N_{\rm H}$ distribution into
the observed one, similar to that used by \cite{ueda2003}. Here we
consider the X-ray luminosity function (XLF), the co-moving number
density per unit co-moving volume per log $L_{\rm X}$ interval as a
function of $L_{\rm X}$ and $z$, to predict the expected number of AGNs
with a given range of $N_{\rm H}$ from the SXDS survey,
using the area curve given as a function of count rate
\citep{ueda2008}. Then, by simulating the statistical fluctuation in the
observed count rates (and hence hardness ratios) according to the real
survey characteristics, we reproduce the expected distribution of the
best-fit $N_{\rm H}$ from each input $N_{\rm H}$ value. Finally, we
perform a fit to the observed histogram of $N_{\rm H}$ with the model
prediction calculated via the matrix, by changing the type-2 fraction as
a free parameter. With this procedure, we can most
reliably constrain the ``intrinsic'' absorption distribution in a {\it
statistical} sense, even if it is hard to determine $N_{\rm H}$ of an
individual object.

For the shape of the 2--10 keV XLF, we adopt the
luminosity-dependent density evolution (LDDE) model of \cite{ueda2003}
by further introducing a power-law decline of the comoving space density
at high redshifts as reported in subsection~\ref{subsec:space
density}. Thus, the evolution factor in equation~(16) of \cite{ueda2003}
is replaced by the form:
%%%\tiny
\begin{equation}
  e(z, L_{\rm X}) = \left\{ \begin{array}{ll}
    \displaystyle (1 + z)^{p1} & [z < z_{c}(L_{\rm X})] \\
    \displaystyle (1 + z_{c})^{p1} \left(\frac{\displaystyle 1 + z}{\displaystyle 1 + z_{c}}\right)^{p2} & [z_{c}(L_{\rm X}) \le z < z_{c2}] \\
    \displaystyle (1 + z_{c})^{p1} \left(\frac{\displaystyle 1 + z_{c2}}{\displaystyle 1 + z_{c}}\right)^{p2} \left(\frac{\displaystyle 1 + z}{\displaystyle 1 + z_{c2}}\right)^{p3} & [z \ge z_{c2}], \\
  \end{array} \right.
\end{equation}
%%%\normalsize
where $p1 = 4.23$, $p2 = -1.5$, and $z_{c} = 1.9$ are adopted, following \cite{ueda2003}.
Here the cutoff redshift, beyond which the comoving space density of AGNs
decreases, is set to $z_{c2} = 2.7$, following \cite{schmidt1995}. We adopt
$p3 = -6.2$ for the power-law slope at $z > z_{c2}$ based on our result
from both the SXDS and COSMOS samples.  We also use the ``$N_{\rm H}$
function'' introduced by \cite{ueda2003}, $f(N_{\rm H})$, i.e. the normalized
probability distribution function for the absorption column density. In
our calculation, $f(N_{\rm H})$ is assumed to be independent of
luminosity and redshift for simplicity. 
We confirm that the effect caused by the luminosity
dependence of the type-2 fraction \citep[e.g.,][]{ueda2003, hasinger2008} on our
result is smaller than the statistical
uncertainty, because the luminosity range of our sample is narrow (log $L_{\rm X}$ = 44--45).

\begin{figure}
  \epsscale{1.2}
  \plotone{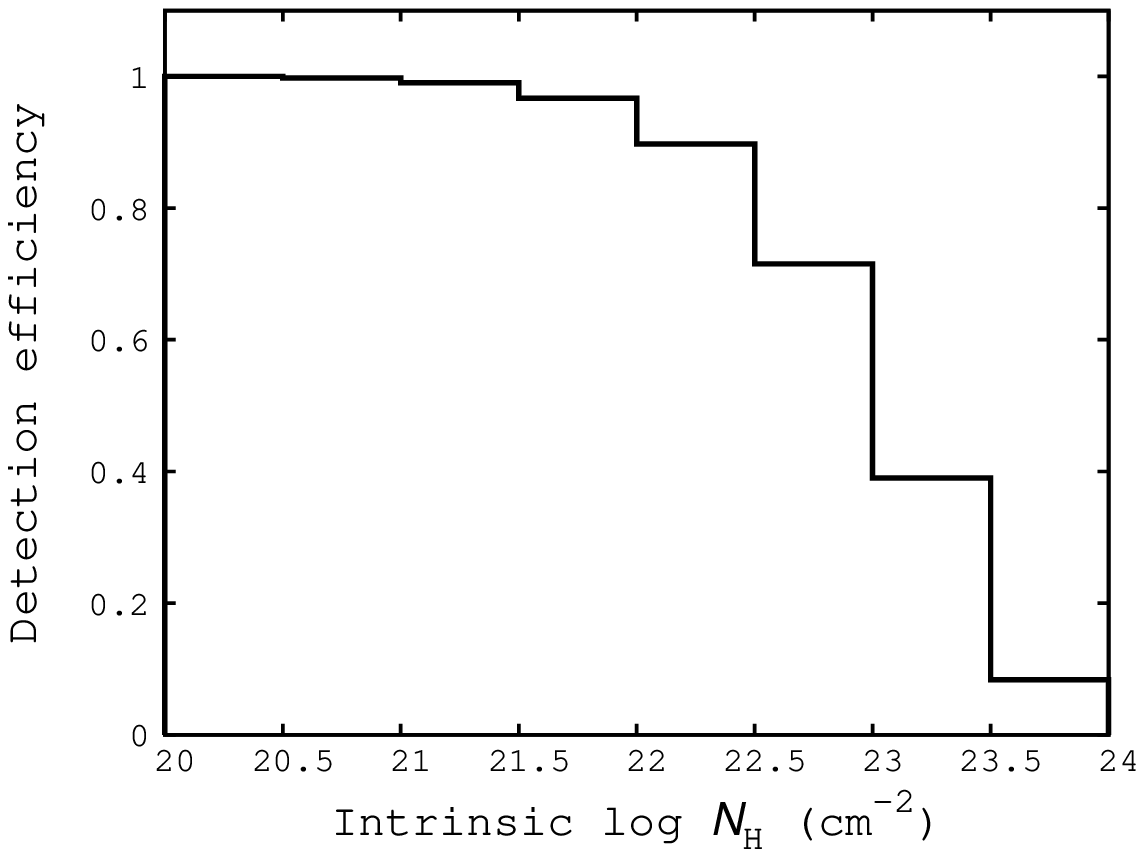}
  \caption{Detection efficiency as a function of intrinsic absorption column
    density, normalized to unity at log $N_{\rm H}$ = 20.0--20.5.
 \label{fig:detection-efficiency}}
\end{figure}

The expected number of AGNs detected, $N$, in the intrinsic
luminosity range log $L_{\rm X}$ = 44--45, $z$ = 3--5, and intrinsic column
density range  $N_{\rm H}$ = $N_{\rm H, min}$--$N_{\rm H, max}$ is 
computed as
%%%\scriptsize
\begin{eqnarray}
  \lefteqn{N = \int^{N_{\rm H,max}}_{N_{\rm H, min}} d {\rm log} N_{\rm H} \int^{45}_{44} d {\rm log} L_{\rm X} \int^{5}_{3} d z} \nonumber \\
  & & f(N_{\rm H}) \frac{d \Phi(L_{\rm X}, z)}{d {\rm log} L_{\rm X}} \Omega(c(L_{\rm X}, z, N_{\rm H})) (1 + z)^{3} d_{A}(z)^{2} c\frac{d\tau}{dz}(z). \nonumber \\
  & &
\end{eqnarray}
%%%\normalsize
The $N$ values, calculated by assuming a flat $N_{\rm H}$
function, give the relative detection efficiency for AGNs with
different column densities. Figure~\ref{fig:detection-efficiency}
shows the result, normalized to unity at log $N_{\rm H}$ = 20.0--20.5.

%%%%%%%%%%%%%%%%%
%   Figure 10   %
%%%%%%%%%%%%%%%%%
%Fig.10

\begin{figure}
  \epsscale{1.2}
  \plotone{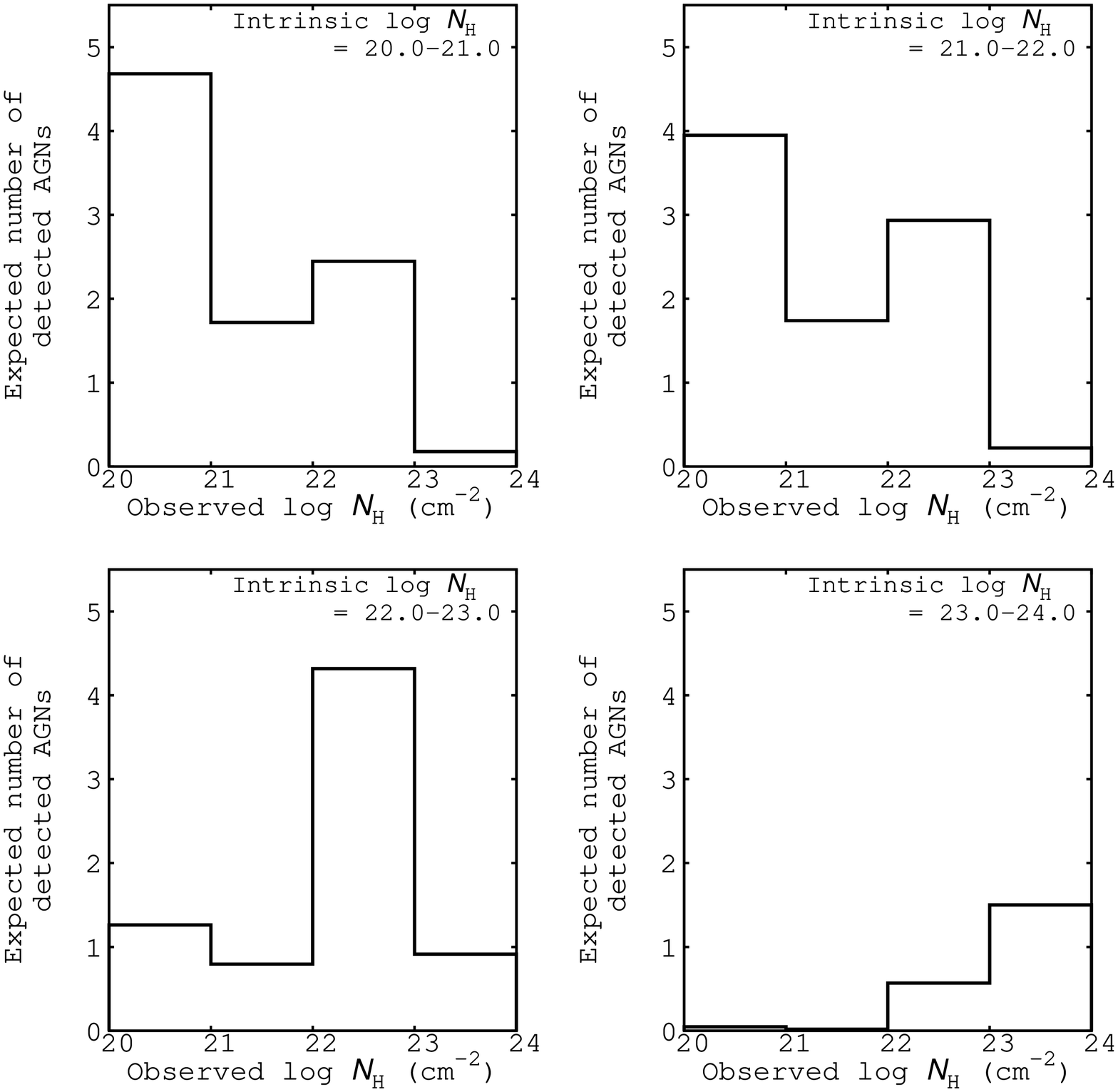}
  \caption{Conversion matrix from the intrinsic $N_{\rm H}$ distribution
    into the observed one. The assumed intrinsic $N_{\rm H}$ distribution
    is plotted in the top right of each panel.
    \label{fig:conversion_matrix}}
\end{figure}

Figure~\ref{fig:conversion_matrix} illustrates the conversion matrix
from the intrinsic to observed $N_{\rm H}$ distributions, calculated for
a flat $f(N_{\rm H})$ in the range of log $N_{\rm H}$ = 20--24.
Each panel of figure~\ref{fig:conversion_matrix} represents
the expected $N_{\rm H}$ distribution for AGNs with input 
intrinsic column
densities of log $N_{\rm H}$ = 20--21, log $N_{\rm H}$ = 21--22, 
log $N_{\rm H}$ = 22--23, and log $N_{\rm H}$ = 23--24. 
Here, it is again emphasized that the statistical errors 
in the count rates (eventually, those in the best-fit $N_{\rm H}$ 
values) are taken into account, based on the observed correlation between 
the count rate and its error, from the re-analysis described 
in section~\ref{sec:Absorption Column Density}. 
It is seen from figure~\ref{fig:conversion_matrix} that the total number of
detected AGNs decreases with the input $N_{\rm H}$ and the 
observed $N_{\rm H}$ distribution has a wide dispersion with respect to the 
input $N_{\rm H}$ value due to statistical fluctuations.

%%%%%%%%%%%%%%%%%
%   Figure 11   %
%%%%%%%%%%%%%%%%%
%Fig.11

\begin{figure}
  \epsscale{1.2}
  \plotone{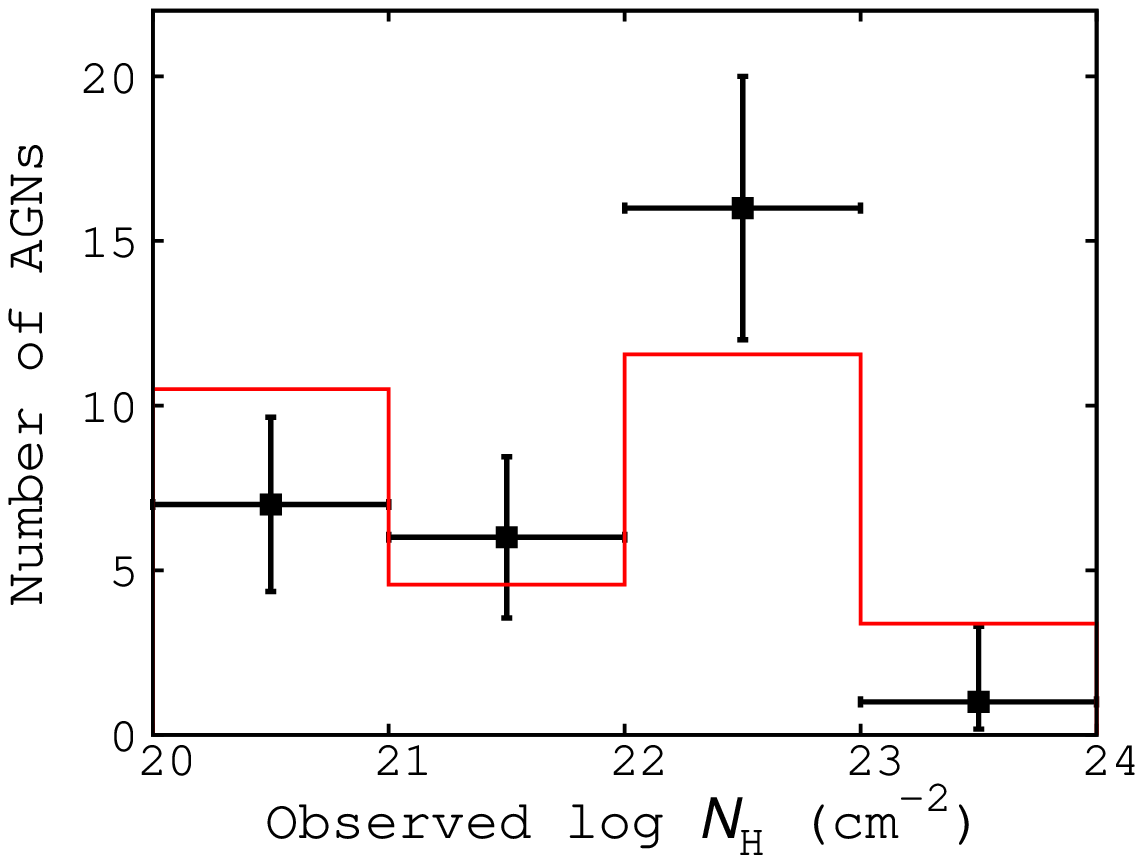}
  \caption{Observed histogram of $N_{\rm H}$ with 1$\sigma$ Poisson errors (black squares).
    The red solid line represents the prediction from the $N_{\rm H}$ function with the best-fit type-2 AGN fraction.
    \label{fig:fitting_result-4bin_norm-fix}}
\end{figure}

Using this matrix, we perform a Poisson maximum likelihood fitting to
the observed $N_{\rm H}$ histogram of our AGN sample. We assume a flat
shape for the $N_{\rm H}$ function in two \nh\ regions, log \nh\ =
20--22 and log \nh\ = 22--24, with the type-2 AGN fraction as a free
parameter to be determined through the fitting. The shape of the XLF
is fixed to the model described above, while its normalization is
tuned to reproduce the total number of observed AGNs at $z$ = 3--5 for
our best-fit \nh\ function. It is confirmed, however, that the choice
of the XLF model does not affect
our conclusion, by introducing the luminosity and
density evolution (LADE) model \citep{aird2010} instead of the LDDE
model. We finally obtain a X-ray type-2 AGN fraction of
0.54$^{+0.17}_{-0.19}$ from our sample.
Figure~\ref{fig:fitting_result-4bin_norm-fix} shows the observed
histogram of $N_{\rm H}$ with 1$\sigma$ Poisson errors (black
squares), on which the model prediction from the \nh\ function with the best-fit type-2
AGN fraction is overplotted (red solid line).

%%%%%%%%%%%%%%%%%
%   Figure 12   %
%%%%%%%%%%%%%%%%%
%Fig.12

\subsubsection{Optical Type-2 Fraction}
\label{sec:opt_f_type2}

To confirm the result obtained in the previous
subsection, here we also estimate the type-2 AGN fraction based on the
optical type of each AGN, not by referring to the X-ray absorption.
In our sample, we have a total of 17 optical type-1 and 13 optical
type-2 AGNs. To derive the intrinsic optical type-2 AGN fraction, we
have to assume a relationship between optical type and absorption column
density, since the latter determines the observational biases in an
X-ray survey. Here, we simply assume that our optical type-1 and type-2
AGNs correspond to those with log \nh\ $< 22$ and log \nh\ = 22--24,
respectively, following \cite{ueda2003}. Then, we can utilize the
detection efficiency plotted in figure~\ref{fig:detection-efficiency} to
correct for the detection bias. Since the optical properties are
independent of the X-ray spectral information, we do
not need to consider the statistical error in \nh\ in contrast with the
previous case. In this way, the optical type-2 fraction for AGNs with
log $L_{\rm X}$ = 44--45 erg s$^{-1}$ at $z$ = 3--5 is estimated to be
$f_{\rm type-2}$ = 0.59$\pm$0.09. 
Even when we consider the possible uncertainties in the classification
of the 10 photometric redshift sample (section~\ref{sec:Classification of AGNs}), 
the type-2 AGN fraction can be at least 0.52$\pm$0.09. These values 
are consistent with the X-ray result. The statistical error becomes
smaller than that for the X-ray type-2 fraction because there is no
coupling between the two types caused by the statistical fluctuation
in the $N_{\rm H}$ value. We must note, however, that the
correspondence of the optical type to X-ray absorption may have some
ambiguities that could produce additional systematic uncertainties.

\subsubsection{Evolution of Type-2 Fraction in Luminous AGNs}

By fully taking into account the observational biases, including the
photon statistics of the X-ray survey data, we have derived the X-ray
type-2 AGN (log \nh\ $>$ 22) fraction in the Compton-thin population to
be 0.54$^{+0.17}_{-0.19}$ at $z$ = 3--5 and in the luminosity range of
log $L_{\rm X}$ = 44--45. The consistency with the
optical type-2 fraction of 0.59$\pm$0.09, which is independently
measured, supports the robustness of our result. 
Recently, \cite{masters2012} roughly estimate the 
type-2 AGN fraction with log $L_{\rm X} > 44.15$ at $z \sim$3--5 to 
be $\sim$75\%, by comparing the normalizations of X-ray (type-1 + 
type-2) and UV (type-1 only) luminosity functions derived from the 
COSMOS data. Their estimate is somewhat larger than our result, 
although it should be very sensitive to the adopted $\alpha_{ox}$ 
value used for the luminosity conversion.

We can now compare our X-ray type-2 AGN fraction with 
that obtained at lower redshifts to investigate its cosmological evolution. 
Recently, \cite{burlon2011} have compiled an AGN
sample in the local Universe detected in the 3-year Swift/BAT survey
performed in the 15--55 keV band, where no detection bias is expected
for Compton-thin AGNs (log \nh\ $< 24$). We find the X-ray type-2
fraction in their sample to be $0.22\pm0.06$ in the luminosity range as
corresponding to log $L_{\rm X}$ = 44--45 in the 2--10
keV band, converted assuming a photon index of 2.  Our result at $z >
3$ is significantly higher than the local value, at a
$\approx$90\% confidence level based on a $\chi^2$ test, by a factor of
$2.5\pm1.1$ (the error is $1\sigma$). When we instead adopt the optical
classification for the SXDS sample, the factor becomes $2.7\pm0.9$ and
the confidence level corresponds to $99.9$\%. Thus, we establish the
evolution of type-2 fraction from $z = 0$ to $z > 3$ in the
high-luminosity range, where it has not been well constrained so far.

We confirm that our results are consistent with the formulation of
\cite{hasinger2008} for the type-2 fraction given as a
function of luminosity and redshift. He finds that the type-2 fraction
normalized at log \lx\ = 43.75 (to correct for the luminosity
dependence) increases as $(1+z)^{0.62}$ and then saturates above $z
\simeq 2$. The formula given in \cite{hasinger2008} gives
%%%\scriptsize
\begin{eqnarray}
  f_{\rm type-2} = (-0.11\pm0.12) * ({\rm log}~L_{\rm X} - 43.75) + (0.578\pm0.081) \nonumber \\
  (z = 3.2-5.2).
\end{eqnarray}
%%%\normalsize
By substituting log $L_{\rm X}$ = 44.6 as the averaged luminosity in our
sample, the type-2 fraction is calculated to be
0.49$\pm$0.13 at $z$ = 3.2--5.2.  Note that Hasinger's
``type-2'' definition utilizes both optical and X-ray information, which
is not exactly the same as ours; in fact, he argues that the criterion
roughly corresponds to log \nh\ = 21.5, instead of 22, and therefore we
have to reduce $f_{\rm type-2}$ to match our definition of X-ray type-2
fraction. If we assume a flat \nh\ function above log \nh\ = 21.5, the
value of type-2 AGN fraction in his definition is reduced by 20\% in our
definition; specifically, $f_{\rm type-2}=0.49\pm0.13$
in his definition corresponds to an X-ray type-2 fraction of
0.39$\pm$0.10 in ours. Thus, the formula of
\cite{hasinger2008} is in agreement with the X-ray type-2 fraction
derived from our SXDS sample within the errors, although our results may
favor a slightly higher type-2 fraction at $z > 3$.

\section{Conclusions}
\label{sec:SUMMARY}

Thanks to the deep and wide area coverage of the SXDS survey with extensive
multi-wavelength follow-up observations, we are able to construct a
highly complete X-ray-selected AGN sample at high redshifts ($3 < z <
5$), consisting of 20 and 10 AGNs with spectroscopic and photometric redshifts, 
respectively. The conclusions are summarized as follows:

\begin{itemize}

\item We derive the comoving space density of Compton-thin AGNs at $z$ =
      3--5 with intrinsic luminosities of $L_{\rm X}$ = 10$^{44-45}$ erg
      s$^{-1}$ at $z > 3$, using the $1/V_{\rm max}$ method. We confirm
      the trend that the space density significantly declines with
      redshift as reported in previous works. Combining our SXDS result
      with that from the COSMOS survey \citep{civano2011}, we find that
      the comoving space density decreases as $(1 + z)^{-6.2\pm0.9}$ or
      $10^{-(0.60\pm0.08)\times(z-2.7)}$, providing so-far the
      best constraint on the declining profile of
      X-ray-selected AGNs at $z > 3$.

\item By carefully taking into account observational biases, including
      statistical uncertainties in the absorption column densities
      derived from the X-ray data, we derive the obscured AGN fraction
      with $N_{\rm H}> 10^{22}$ cm$^{-2}$ among those with $N_{\rm H}
      <10^{24}$ cm$^{-2}$ to be 0.54$^{+0.17}_{-0.19}$ at $z$ = 3--5 in
      the luminosity range of $L_{\rm X}$ = 10$^{44-45}$ erg s$^{-1}$.
      From comparison with the results obtained at lower redshifts, we
      establish that there is a significant redshift evolution of obscured
      fraction from $z = 0$ to $z > 3$ by a factor of
      $2.5\pm1.1$ in the high-luminosity range.

\end{itemize}

\acknowledgments

We thank the anonymous referee for the useful comments
that help to improve the clarity of the paper. This work was partly
supported by the Grant-in-Aid for JSPS Fellows for young researchers
(KH), Scientific Research 23540265 (YU), and by the grant-in-aid for the
Global COE Program ``The Next Generation of Physics, Spun from
Universality and Emergence'' from the Ministry of Education, Culture,
Sports, Science and Technology (MEXT) of Japan.

%%%\appendix
%%%
%%%\section{Appendix material}
%%%
%%%APPENDIX

\clearpage
%%%\LongTables % optionally
\begin{landscape}

\begin{deluxetable*}{ccccccccccccccc}
  \tablefontsize{\footnotesize}
  \tabletypesize{\scriptsize}
  \tablecaption{Properties of our high-redshift AGN sample. \label{tab:sample}}
  \tablewidth{0pt}
  \tablehead{
    \colhead{}       & \colhead{R.A.}  & \colhead{Dec.}  & \colhead{Optical}     & \colhead{SED}   & \colhead{}   & \colhead{}   & \colhead{}      & \multicolumn{3}{c}{Count Rate (counts ks$^{-1}$)} & \colhead{} & \colhead{} & \colhead{} & \colhead{} \\
      \cline{9-11}\\[-2mm]
      \colhead{SXDF ID} & \colhead{(deg)} & \colhead{(deg)} & \colhead{Type$^{[a]}$} & \colhead{Model$^{[b]}$} & \colhead{$z^{[c]}$} & \colhead{$z_{\rm spec}^{[c]}$} & \colhead{$z_{\rm phot}^{[c]}$}  & \colhead{0.5--2} & \colhead{0.3--1.0} & \colhead{1.0--4.5} & \colhead{HR$^{[d]}$} & \colhead{log $N_{\rm H}$$^{[e]}$} & \colhead{Flux$^{[f]}$} & \colhead{$L_{\rm X}$$^{[g]}$}
    }
    \startdata
0016 & 33.93335 & -4.92384 & 1 & 1 & 3.512 & 3.512 & 3.298 & 1.40 $\pm$ 0.36 & 0.93 $\pm$ 0.29 & 1.22 $\pm$ 0.34 & 0.13 $\pm$ 0.20 &	$22.30_{-2.30}^{+0.41}$ &	3.16 & 3.57 \\
0099 & 34.08598 & -5.28820 & 1 & 1 & 3.190 & 3.190 & 3.154 & 2.38 $\pm$ 0.47 & 3.16 $\pm$ 0.51 & 1.89 $\pm$ 0.38 & -0.25 $\pm$ 0.12 &	$20.00_{-0.00}^{+0.00}$ &	4.68 & 4.19 \\
0154 & 34.14188 & -4.90640 & 0 & 2 & 3.600 & ... & 3.600 & 0.64 $\pm$ 0.20 & 0.39 $\pm$ 0.19 & 0.63 $\pm$ 0.20 & 0.23 $\pm$ 0.27 &	$22.55_{-1.27}^{+0.44}$ &	1.58 & 1.89 \\
0177 & 34.16076 & -5.17883 & 1 & 2 & 3.182 & 3.182 & 3.282 & 1.66 $\pm$ 0.33 & 0.96 $\pm$ 0.32 & 0.90 $\pm$ 0.34 & -0.03 $\pm$ 0.25 &	$21.44_{-1.44}^{+0.99}$ &	3.35 & 2.98 \\
0179 & 34.16469 & -4.72107 & 0 & 2 & 3.426 & ... & 3.426 & 0.63 $\pm$ 0.20 & 1.34 $\pm$ 0.25 & 0.40 $\pm$ 0.20 & -0.54 $\pm$ 0.19 &	$20.00_{-0.00}^{+0.00}$ &	1.23 & 1.31 \\
0284 & 34.24511 & -4.81274 & 0 & 2 & 3.046 & ... & 3.046 & 1.22 $\pm$ 0.30 & 0.87 $\pm$ 0.28 & 0.79 $\pm$ 0.27 & -0.05 $\pm$ 0.24 &	$21.10_{-1.10}^{+1.24}$ &	2.44 & 1.96 \\
0287 & 34.24634 & -4.83036 & 0 & 2 & 4.090 & ... & 4.090 & 2.55 $\pm$ 0.42 & 1.60 $\pm$ 0.36 & 1.65 $\pm$ 0.37 & 0.02 $\pm$ 0.16 &	$21.98_{-1.98}^{+0.57}$ &	5.12 & 8.30 \\
0335 & 34.27461 & -5.22714 & 2 & 2 & 3.222 & 3.222 & 3.054 & 0.76 $\pm$ 0.18 & 0.26 $\pm$ 0.14 & 0.75 $\pm$ 0.14 & 0.49 $\pm$ 0.22 &	$22.88_{-0.34}^{+0.31}$ &	2.60 & 2.39 \\
0342 & 34.28106 & -4.56844 & 0 & 2 & 3.048 & ... & 3.048 & 0.88 $\pm$ 0.18 & 0.09 $\pm$ 0.10 & 1.11 $\pm$ 0.19 & 0.86 $\pm$ 0.16 &	$23.40_{-0.27}^{+0.60}$ &	6.57 & 5.28 \\
0385 & 34.32280 & -5.08699 & 0 & 1 & 3.390 & ... & 3.390 & 1.13 $\pm$ 0.25 & 0.58 $\pm$ 0.21 & 0.83 $\pm$ 0.23 & 0.17 $\pm$ 0.22 &	$22.38_{-1.33}^{+0.40}$ &	2.66 & 2.76 \\
0422 & 34.34430 & -5.39438 & 2 & 2 & 3.422 & 3.422 & 1.840 & 2.46 $\pm$ 0.15 & 1.52 $\pm$ 0.21 & 2.33 $\pm$ 0.25 & 0.21 $\pm$ 0.08 &	$22.47_{-0.19}^{+0.15}$ &	6.02 & 6.38 \\
0449 & 34.35595 & -4.98441 & 1 & 1 & 3.328 & 3.328 & 3.038 & 1.13 $\pm$ 0.19 & 0.78 $\pm$ 0.15 & 1.02 $\pm$ 0.18 & 0.13 $\pm$ 0.13 &	$22.26_{-0.53}^{+0.28}$ &	2.58 & 2.56 \\
0459 & 34.36505 & -5.28871 & 1 & 1 & 3.969 & 3.969 & 3.946 & 0.84 $\pm$ 0.17 & 0.71 $\pm$ 0.13 & 0.55 $\pm$ 0.12 & -0.13 $\pm$ 0.14 &	$20.00_{-0.00}^{+1.89}$ &	1.60 & 2.42 \\
0489 & 34.38058 & -5.09214 & 0 & 2 & 3.334 & ... & 3.334 & 1.43 $\pm$ 0.22 & 0.61 $\pm$ 0.15 & 1.65 $\pm$ 0.23 & 0.46 $\pm$ 0.11 &	$22.86_{-0.16}^{+0.16}$ &	4.65 & 4.63 \\
0508 & 34.39336 & -5.08771 & 1 & 1 & 3.975 & 3.975 & 4.164 & 0.98 $\pm$ 0.18 & 0.71 $\pm$ 0.14 & 1.02 $\pm$ 0.18 & 0.18 $\pm$ 0.13 &	$22.53_{-0.41}^{+0.26}$ &	2.27 & 3.45 \\
0520 & 34.39922 & -4.70872 & 1 & 1 & 3.292 & 3.292 & 2.988 & 0.58 $\pm$ 0.17 & 0.27 $\pm$ 0.14 & 0.51 $\pm$ 0.17 & 0.31 $\pm$ 0.28 &	$22.62_{-0.72}^{+0.42}$ &	1.57 & 1.52 \\
0564 & 34.43107 & -4.54404 & 2 & 2 & 3.204 & 3.204 & 2.944 & 1.25 $\pm$ 0.27 & 0.78 $\pm$ 0.23 & 1.12 $\pm$ 0.23 & 0.18 $\pm$ 0.17 &	$22.36_{-0.58}^{+0.31}$ &	3.01 & 2.72 \\
0650 & 34.49477 & -5.13762 & 1 & 1 & 3.032 & 3.032 & 2.708 & 1.10 $\pm$ 0.14 & 0.87 $\pm$ 0.15 & 0.71 $\pm$ 0.16 & -0.10 $\pm$ 0.14 &	$20.00_{-0.00}^{+1.90}$ &	2.18 & 1.73 \\
0700 & 34.52910 & -5.06942 & 1 & 2 & 3.128 & 3.128 & 2.986 & 2.39 $\pm$ 0.09 & 1.02 $\pm$ 0.14 & 2.34 $\pm$ 0.19 & 0.39 $\pm$ 0.07 &	$22.71_{-0.11}^{+0.10}$ &	7.22 & 6.18 \\
0742 & 34.56485 & -5.40081 & 1 & 1 & 3.114 & 3.114 & 1.360 & 3.30 $\pm$ 0.43 & 2.55 $\pm$ 0.38 & 2.53 $\pm$ 0.37 & -0.00 $\pm$ 0.11 &	$21.70_{-1.70}^{+0.46}$ &	6.84 & 5.78 \\
0788 & 34.59990 & -5.10029 & 0 & 2 & 4.096 & ... & 4.096 & 1.00 $\pm$ 0.13 & 0.63 $\pm$ 0.13 & 0.71 $\pm$ 0.16 & 0.06 $\pm$ 0.15 &	$22.19_{-2.19}^{+0.43}$ &	2.07 & 3.37 \\
0809 & 34.61807 & -5.26411 & 1 & 2 & 3.857 & 3.857 & 3.504 & 0.98 $\pm$ 0.23 & 0.66 $\pm$ 0.21 & 0.66 $\pm$ 0.21 & -0.00 $\pm$ 0.22 &	$21.81_{-1.81}^{+0.78}$ &	1.96 & 2.76 \\
0824 & 34.63105 & -4.73291 & 1 & 1 & 3.699 & 3.699 & 3.766 & 1.15 $\pm$ 0.35 & 0.50 $\pm$ 0.26 & 0.87 $\pm$ 0.36 & 0.27 $\pm$ 0.31 &	$22.65_{-1.38}^{+0.48}$ &	2.95 & 3.77 \\
0835 & 34.64068 & -5.28748 & 1 & 1 & 3.553 & 3.553 & 3.210 & 0.69 $\pm$ 0.18 & 0.48 $\pm$ 0.16 & 0.38 $\pm$ 0.14 & -0.11 $\pm$ 0.24 &	$20.00_{-0.00}^{+2.31}$ &	1.34 & 1.55 \\
0888 & 34.68521 & -4.80691 & 1 & 1 & 4.550 & 4.550 & 3.598 & 1.66 $\pm$ 0.43 & 0.97 $\pm$ 0.38 & 1.35 $\pm$ 0.43 & 0.17 $\pm$ 0.25 &	$22.62_{-2.62}^{+0.47}$ &	3.72 & 7.76 \\
0904 & 34.69847 & -5.38866 & 1 & 1 & 3.020 & 3.020 & 2.852 & 1.09 $\pm$ 0.20 & 1.05 $\pm$ 0.17 & 0.55 $\pm$ 0.15 & -0.31 $\pm$ 0.14 &	$20.00_{-0.00}^{+0.00}$ &	2.16 & 1.69 \\
0926 & 34.72080 & -5.01810 & 2 & 1 & 3.264 & 3.264 & 1.894 & 2.91 $\pm$ 0.47 & 0.79 $\pm$ 0.35 & 2.80 $\pm$ 0.42 & 0.56 $\pm$ 0.16 &	$22.99_{-0.23}^{+0.23}$ &	10.90 & 10.40 \\
0930 & 34.72563 & -5.52342 & 0 & 2 & 3.490 & ... & 3.490 & 2.28 $\pm$ 0.41 & 1.35 $\pm$ 0.35 & 1.37 $\pm$ 0.29 & 0.01 $\pm$ 0.17 &	$21.83_{-1.83}^{+0.59}$ &	4.67 & 5.19 \\
1032 & 34.80911 & -5.17238 & 0 & 2 & 3.584 & ... & 3.584 & 1.91 $\pm$ 0.39 & 1.28 $\pm$ 0.34 & 1.56 $\pm$ 0.32 & 0.10 $\pm$ 0.17 &	$22.23_{-2.23}^{+0.40}$ &	4.19 & 4.96 \\
1238 & 35.09164 & -5.07500 & 1 & 2 & 4.174 & 4.174 & 3.466 & 1.37 $\pm$ 0.35 & 1.25 $\pm$ 0.32 & 0.89 $\pm$ 0.32 & -0.17 $\pm$ 0.21 &	$20.00_{-0.00}^{+2.10}$ &	2.60 & 4.42 \\
    \enddata
    \tablenotetext{a}{0: photometric redshift AGNs, 1: optical type-1 AGNs, and 2: optical type-2 AGNs}
    \tablenotetext{b}{SED models used for the calculations of photometric redshifts: 1; QSO template and 2; Galaxy template}
    \tablenotetext{c}{Chosen redshift; spectroscopic redshift; photometric redshift}
    \tablenotetext{d}{Hardness ratio defined as ``HR = $(H+S)/(H-S)$'', S: 0.3--1.0 keV count rate and H: 1.0--4.5 keV count rate}
    \tablenotetext{e}{The absorption column density in units of cm$^{-2}$.}
    \tablenotetext{f}{The intrinsic (de-absorbed and rest-frame 2--10 keV) flux in units of 10$^{-15}$ erg cm$^{-2}$ s$^{-1}$.}
    \tablenotetext{g}{The intrinsic (de-absorbed and rest-frame 2--10 keV) luminosity in units of 10$^{44}$ erg s$^{-1}$.}
\end{deluxetable*}

\clearpage
\end{landscape}


\begin{thebibliography}{}

\bibitem[Aird et al.(2010)]{aird2010} 
  Aird, J., Nandra, K., Laird, E.~S., et al.\ 2010, \mnras, 401, 2531 

\bibitem[Akiyama et al.(2000)]{akiyama2000} 
  Akiyama, M., Ohta, K., Yamada, T., et al.\ 2000, \apj, 532, 700 

\bibitem[Antonucci(1993)]{antonucci1993} 
  Antonucci, R.\ 1993, \araa, 31, 473 
  
\bibitem[Barger et al.(2005)]{barger2005} 
  Barger, A.~J., Cowie, L.~L., Mushotzky, R.~F., et al.\ 2005, \aj, 129, 578 

\bibitem[Bolzonella et al.(2000)]{bolzonella2000} 
  Bolzonella, M., Miralles, J.-M., \& Pell{\'o}, R.\ 2000, \aap, 363, 476 
  
\bibitem[Brammer et al.(2008)]{brammer2008} 
  Brammer, G.~B., van Dokkum, P.~G., \& Coppi, P.\ 2008, \apj, 686, 1503 

\bibitem[Brusa et al.(2009)]{brusa2009} 
  Brusa, M., Comastri, A., Gilli, R., et al.\ 2009, \apj, 693, 8 

\bibitem[Burlon et al.(2011)]{burlon2011} 
  Burlon, D., Ajello, M., Greiner, J., et al.\ 2011, \apj, 728, 58 
  
\bibitem[Cardamone et al.(2010)]{cardamone2010} 
  Cardamone, C.~N., van Dokkum, P.~G., Urry, C.~M., et al.\ 2010, \apjs, 189, 270 
    
\bibitem[Civano et al.(2011)]{civano2011} 
  Civano, F., Brusa, M., Comastri, A., et al.\ 2011, \apj, 741, 91 

\bibitem[Cowie et al.(1996)]{cowie1996} 
  Cowie, L.~L., Songaila, A., Hu, E.~M., \& Cohen, J.~G.\ 1996, \aj, 112, 839 
  
\bibitem[Ebrero et al.(2009)]{ebrero2009} 
  Ebrero, J., Carrera, F.~J., Page, M.~J., et al.\ 2009, \aap, 493, 55 

\bibitem[Elvis et al.(2009)]{elvis2009} 
  Elvis, M., Civano, F., Vignali, C., et al.\ 2009, \apjs, 184, 158 
  
\bibitem[Ferrarese \& Merritt(2000)]{ferrarese2000} 
  Ferrarese, L., \& Merritt, D.\ 2000, \apjl, 539, L9 

\bibitem[Furusawa et al.(2008)]{furusawa2008} 
  Furusawa, H., Kosugi, G., Akiyama, M., et al.\ 2008, \apjs, 176, 1 

\bibitem[Gebhardt et al.(2000)]{gebhardt2000} 
  Gebhardt, K., Bender, R., Bower, G., et al.\ 2000, \apjl, 539, L13 

\bibitem[Gehrels(1986)]{gehrels1986} 
  Gehrels, N.\ 1986, \apj, 303, 336 

\bibitem[Gilli et al.(2007)]{gilli2007} 
  Gilli, R., Comastri, A., \& Hasinger, G.\ 2007, \aap, 463, 79 
  
\bibitem[Glikman et al.(2011)]{glikman2011} 
  Glikman, E., Djorgovski, S.~G., Stern, D., et al.\ 2011, \apjl, 728, L26 

\bibitem[Hasinger et al.(2005)]{hasinger2005} 
  Hasinger, G., Miyaji, T., \& Schmidt, M.\ 2005, \aap, 441, 417 
  
\bibitem[Hasinger et al.(2007)]{hasinger2007} 
  Hasinger, G., Cappelluti, N., Brunner, H., et al.\ 2007, \apjs, 172, 29 

\bibitem[Hasinger(2008)]{hasinger2008} 
  Hasinger, G.\ 2008, \aap, 490, 905 

\bibitem[Ikeda et al.(2011)]{ikeda2011} 
  Ikeda, H., Nagao, T., Matsuoka, K., et al.\ 2011, \apjl, 728, L25 

\bibitem[Jiang et al.(2009)]{jiang2009} 
  Jiang, L., Fan, X., Bian, F., et al.\ 2009, \aj, 138, 305 

\bibitem[Kodama et al.(2004)]{kodama2004} 
  Kodama, T., Yamada, T., Akiyama, M., et al.\ 2004, \mnras, 350, 1005 

\bibitem[La Franca et al.(2005)]{lafranca2005} 
  La Franca, F., Fiore, F., Comastri, A., et al.\ 2005, \apj, 635, 864 

\bibitem[Luo et al.(2010)]{luo2010} 
  Luo, B., Brandt, W.~N., Xue, Y.~Q., et al.\ 2010, \apjs, 187, 560 

\bibitem[Magorrian et al.(1998)]{magorrian1998} 
  Magorrian, J., et al.\ 1998, \aj, 115, 2285 

\bibitem[Masters et al.(2012)]{masters2012} 
  Masters, D., Capak, P., Salvato, M., et al.\ 2012, arXiv:1207.2154 
  
\bibitem[Mortlock et al.(2011)]{mortlock2011} 
  Mortlock, D.~J., Warren, S.~J., Venemans, B.~P., et al.\ 2011, \nat, 474, 616 

\bibitem[Richards et al.(2006)]{richards2006} 
  Richards, G.~T., Strauss, M.~A., Fan, X., et al.\ 2006, \aj, 131, 2766 
  
\bibitem[Schmidt(1968)]{schmidt1968} 
  Schmidt, M.\ 1968, \apj, 151, 393 

\bibitem[Schmidt et al.(1995)]{schmidt1995} 
  Schmidt, M., Schneider, D.~P., \& Gunn, J.~E.\ 1995, \aj, 110, 68 

\bibitem[Scoville et al.(2007)]{scoville2007} 
  Scoville, N., Aussel, H., Brusa, M., et al.\ 2007, \apjs, 172, 1 

\bibitem[Silverman et al.(2008)]{silverman2008} 
  Silverman, J.~D., Green, P.~J., Barkhouse, W.~A., et al.\ 2008, \apj, 679, 118 
\bibitem[Steffen et al.(2003)]{steffen2003} 
  Steffen, A.~T., Barger, A.~J., Cowie, L.~L., Mushotzky, R.~F., \& Yang, Y.\ 2003, \apjl, 596, L23 

\bibitem[Stern \& Laor(2012)]{stern2012} 
  Stern, J., \& Laor, A.\ 2012, \mnras, 423, 600 
  
\bibitem[Str{\"u}der et al.(2001)]{struder2001} 
  Str{\"u}der, L., Briel, U., Dennerl, K., et al.\ 2001, \aap, 365, L18 

\bibitem[Treister \& Urry(2006)]{treister2006} 
  Treister, E., \& Urry, C.~M.\ 2006, \apjl, 652, L79 

\bibitem[Turner et al.(2001)]{turner2001} 
  Turner, M.~J.~L., Abbey, A., Arnaud, M., et al.\ 2001, \aap, 365, L27 
  
\bibitem[Ueda et al.(2003)]{ueda2003} 
  Ueda, Y., Akiyama, M., Ohta, K., \& Miyaji, T.\ 2003, \apj, 598, 886 

\bibitem[Ueda et al.(2008)]{ueda2008} 
  Ueda, Y., Watson, M.~G., Stewart, I.~M., et al.\ 2008, \apjs, 179, 124 

\bibitem[Ueda et al.(2011)]{ueda2011} 
  Ueda, Y., Hiroi, K., Isobe, N., et al.\ 2011, \pasj, 63, 937 

\bibitem[Urry \& Padovani(1995)]{urry1995} 
  Urry, C.~M., \& Padovani, P.\ 1995, \pasp, 107, 803 

\bibitem[Willott et al.(2010)]{willott2010} 
  Willott, C.~J., Delorme, P., Reyl{\'e}, C., et al.\ 2010, \aj, 139, 906 

\bibitem[Yencho et al.(2009)]{yencho2009} 
  Yencho, B., Barger, A.~J., Trouille, L., \& Winter, L.~M.\ 2009, \apj, 698, 380 

\end{thebibliography}
\end{document}